\documentclass[prb,aps,twocolumn,amsfonts,showpacs,eqsecnum,floatfix]{revtex4}

\usepackage{amsmath}
\usepackage{epsfig}
\usepackage{psfrag}
\usepackage{color}
\usepackage{graphicx}
\usepackage{subfigure}

\begin{document}


\title{Transport Properties of Multiple Quantum Dots Arranged in Parallel: Results from the Bethe Ansatz}
\author{Robert M. Konik}
\affiliation{Department of Physics, Brookhaven National Laboratory, Upton,
NY 11973}

\date{\today}

\begin{abstract}
In this paper we analyze transport through a double dot system connected
to two external leads.  Imagining each dot possessing a single active
level, we model the system through a generalization of the Anderson model.
We argue that this model is exactly solvable when certain constraints
are placed upon the dot Coulomb charging energy, the dot-lead hybridization,
and the value of the applied gate voltage.  Using this exact solvability, we
access the zero temperature linear response conductance both in and out of the presence of
a Zeeman field.  We are also able to study the finite temperature linear response conductance.
We focus on universal behaviour and identify three primary features in the transport of the dots: i) a so-called RKKY
Kondo effect; ii) a standard Kondo effect; and iii) interference phenomena leading to sharp
variations in the conductance including conductance zeros.  We are able to use the exact solvability
of the dot model to characterize these phenomena quantitatively.
While here
we primarily consider a double dot system, the approach adopted applies equally well to
N-dot systems.
\end{abstract}
\maketitle

\newcommand{\epo}{\epsilon_{d1}}
\newcommand{\ept}{\epsilon_{d2}}
\newcommand{\tks}{T^{\rm Standard}_K}
\newcommand{\ttk}{\tilde T^{\rm RKKY}_K}
\newcommand{\tk}{T^{\rm RKKY}_K}
\newcommand{\delep}{|\epsilon_{d1}-\epsilon_{d2}|}
\newcommand{\htk}{H/\tilde T^{\rm RKKY}_K}
\newcommand{\del}{\partial}
\newcommand{\ep}{\epsilon}
\newcommand{\clsd}{c_{l\sig}^\dagger}
\newcommand{\cls}{c_{l\sig}}
\newcommand{\cesd}{c_{e\sig}^\dagger}
\newcommand{\ces}{c_{e\sig}}
\newcommand{\up}{\uparrow}
\newcommand{\down}{\downarrow}
\newcommand{\il}{\int^{\tilde{Q}}_Q d\la~}
\newcommand{\ilp}{\int^{\tilde{Q}}_Q d\la '}
\newcommand{\ik}{\int^{B}_{-D} dk~}
\newcommand{\ila}{\int d\la~}
\newcommand{\ilpa}{\int d\la '}
\newcommand{\ika}{\int dk~}
\newcommand{\tQ}{\tilde{Q}}
\newcommand{\rh}{\rho_{\rm bulk}}
\newcommand{\ri}{\rho^{\rm imp}}
\newcommand{\sh}{\sig_{\rm bulk}}
\newcommand{\si}{\sig^{\rm imp}}
\newcommand{\rph}{\rho_{p/h}}
\newcommand{\sph}{\sig_{p/h}}
\newcommand{\rp}{\rho_{p}}
\newcommand{\sip}{\sig_{p}}
\newcommand{\drph}{\delta\rho_{p/h}}
\newcommand{\dsph}{\delta\sig_{p/h}}
\newcommand{\drp}{\delta\rho_{p}}
\newcommand{\dsp}{\delta\sig_{p}}
\newcommand{\drh}{\delta\rho_{h}}
\newcommand{\dsh}{\delta\sig_{h}}
\newcommand{\enp}{\ep^+}
\newcommand{\enm}{\ep^-}
\newcommand{\enpm}{\ep^\pm}
\newcommand{\enph}{\ep^+_{\rm bulk}}
\newcommand{\enmh}{\ep^-_{\rm bulk}}
\newcommand{\enpi}{\ep^+_{\rm imp}}
\newcommand{\enmi}{\ep^-_{\rm imp}}
\newcommand{\enh}{\ep_{\rm bulk}}
\newcommand{\eni}{\ep_{\rm imp}}
\newcommand{\sig}{\sigma}
\newcommand{\la}{\lambda}
\newcommand{\ua}{\uparrow}
\newcommand{\da}{\downarrow}
\newcommand{\ed}{\epsilon_d}
\newcommand{\om}{\omega}

\section{Introduction}

Quantum dot devices exhibit a wide range of strongly correlated phenomena.
In the simplest of cases, single dot devices exhibit classical Kondo physics,\cite{gold,kondo}
the archetype of strong correlations in impurity systems.
This physics arises in dots due to their natural localization of electronic (and hence spin)
degrees of freedom.  Measuring transport properties provides
the most effective means to probe the Kondo physics in such dots.  The formation of the Abrikosov-Suhl
resonance, a dynamical enhancement in the density of states at the Fermi energy, facilitates transport
through the dot from one lead to the other.  Manipulations of this resonance through adjustments of
temperature, bias, or external fields are readily observed through monitoring the device's conductance.

At the next level of complexity are single level dots where multiple energy levels are relevant
to transport.  Much attention has been
paid to the situation where two levels in a single dot are anomalously close.\cite{eto,pus,pus1,wiel}  In such a case the
dot can be tuned through a singlet-triplet transition.  On the triplet side a two stage
Kondo effect can be realized.\cite{pus1,wiel}  Here two channels of electrons coupled with different strengths
to the dot triplet successively reduce the impurity from $S=1$ to $S=1/2$ to $S=0$ in a two stage process.

Quantum dot devices with multiple active dots offer the same range of
phenomenology as single multi-level dots but with greater versatility as
level spacing and interactions are not precisely controllable within a single dot.
In particular multiple quantum dot devices
offer the opportunity to study non-trivial mixtures of strongly correlated physics and interference
phenomena.  Any Kondo physics that arises in such systems will be strongly affected
by the presence of multiple
tunneling paths afforded by the multiple dots.  Even in the absence of interactions, multiple tunneling paths
lead to richly structured transport behaviour predicated upon interference effects.  

Multiple dot devices come in a number of guises.
In one instance they are manufactured from gated semiconducting heterostructures.\cite{RKKY1,RKKY2,scdots1,scdots2}
An important feature that the semiconducting multi-dots 
share with their single dot cousins is their tunability.  By adjusting
voltages applied to various gates which define the dots, 
it is possible to change fundamental parameters of the dot system
allowing experimentally exploration of all of the dots' various physical regimes.
In a second instance they can be constructed from carbon nanotubes.\cite{cndots1,cndots2}
With appropriate gating, carbon nanotube dots can also be made fully tunable.\cite{cndots1,cndots2}  
Much of the experimental
interest in multiple quantum dots comes from their possible realization as elementary qubits in a quantum
computer.\cite{dec1,scdots2,dec3}  As the spin degrees of freedom on the dot serve as the states of the qubit, the question
of spin decoherence due to coupling with the nuclear background becomes an important question.\cite{dec1,scdots2,dec3,dec4}
Carbon nanotube dots have the advantage that their
spin decoherence times are expected to be longer due to a weaker nuclear background 
in carbon than in GaAs.\cite{cndots1,cndots2}

The promised rich phenomenology and the ability to realize multiple dot devices together have led to
a flurry of theoretical work in this area.  
Combinations of interference
effects and Kondo physics in parallel dots have been studied in many works.\cite{simon2,c4,c1,c5,c7,c9,schiller}
In response to the observations of a competition
between an effective Ruderman-Kittel-Kasuya-Yosida (RKKY) coupling and a Kondo effect in a pair
of double dots in series,\cite{RKKY1,RKKY2} a number of works have studied the physics of RKKY
effects in quantum dots.\cite{glazman,simon,c4}    Techniques
that have been used to study multiple dots include infinite U slave boson 
mean field theory,\cite{grempel,simon2,c4,simon,c1,c5,georges,pos}
numerical renormalization group,\cite{grempel,simon2,c8,c9,dias} exact diagonalization,\cite{c7,c3}
appeal to asymptotic limits,\cite{silvestrov,pus,pus1,glazman}, a
perturbative renormalization group\cite{c10}, and a functional renormalization group \cite{meden,meden1}.

In this work we present a distinct approach to analyzing such systems:  we study such systems under
the rubric of integrability.  Like the numerical renormalization
group it offers exact results but with the advantage of analytic control.  Integrability has long
been known to be able to access {\it thermodynamic} properties of single level dot systems.\cite{wie}  
More recently it has been shown to be
successful in computing equilibrium {\it transport} properties of single level
quantum dots.\cite{long}  In particular, these techniques were able to compute the finite temperature
linear response conductance over a number of decades in temperature in good agreement with the numerical
renormalization group.   The aim of this paper is to apply this approach to multiple quantum dots.
It is an extension and elaboration of work first reported in Ref. (\onlinecite{short}).

To this end, we examine a generalized Anderson model coupling two leads ($l=1,2$) to an array of
dots in parallel:
\begin{eqnarray}\label{eIi}
{\cal H} &=& -i\sum_{l\sigma}\int^\infty_{-\infty} dx c^\dagger_{l\sigma}\partial_xc_{l\sigma}
+ \sum_{\sigma\alpha}V_{l\alpha}(c^\dagger_{l\sigma\alpha}d_{\sigma\alpha} + {\rm h.c}) \cr\cr
&& + \sum_{\sigma\alpha}\epsilon_{d\alpha}n_{\sigma\alpha} + \sum_{\alpha\alpha'} U_{\alpha\alpha'} 
n_{\ua\alpha}n_{\da\alpha}.
\end{eqnarray}
Here the $c_{l\sigma}/d_{\alpha\sigma}$ specify electrons living in the leads and the dots.
$\alpha$ indexes the various levels on the dots in the system.  $V_{l\alpha}$ measures
the tunneling strength between the dot level $\alpha$ and lead $l$.  $U_{\alpha\alpha'}$ characterizes
the Coulombic repulsion between electrons of opposite spin living in levels $\alpha$ and $\alpha'$.
A schematic of this Hamiltonian for two dots, i.e. $\alpha = 1,2$, in given in Figure 1.

\begin{figure}[tbh]
\centerline{\psfig{figure=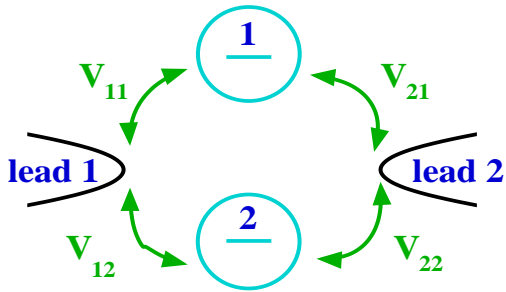,height=1.5in,width=2.6in}}
\caption{A schematic of two dots in parallel.}
\end{figure}

We will argue in this paper that the above Anderson-like Hamiltonian is exactly solvable in two cases:
\begin{eqnarray}\label{eIii}
V_{1\alpha}/V_{2\alpha} &=& V_{1\alpha'}/V_{2\alpha'};\cr\cr
U_{\alpha\alpha '} &=& \delta_{\alpha \alpha'}U_{\alpha};~~U_{\alpha} \Gamma_\alpha = 
U_{\alpha'}\Gamma_{\alpha'};\cr\cr
&& \hskip -.14in U_{\alpha} + 2\epsilon_{d\alpha} = U_{\alpha'} + 2\epsilon_{d\alpha'};
\end{eqnarray}
\vskip 10pt
and 
\begin{eqnarray}\label{eIiii}
V_{1\alpha}/V_{2\alpha} &=& V_{1\alpha'}/V_{2\alpha'};\cr\cr
U_{\alpha\alpha'} &=& U;~~\Gamma_\alpha = \Gamma_{\alpha'};~~\epsilon_{d\alpha} = \epsilon_{d\alpha'} = \epsilon_d.
\end{eqnarray}
The first condition in both cases guarantees that only a single effective channel in the leads 
couples to the dots.  This condition is commonplace in the literature.\cite{c1,c3,c4,c5,c6,c7,c8,c9,c10}.
The second condition implies that for fixed Coulomb repulsion of
the dots, the bare level separation of the two dots, $\delep$, is also fixed.  This still permits
the dot levels to be moved in unison, say by the application of a uniform gate voltage.  
The final condition amounts to the insistence that the energy scale
associated with the charge degrees of freedom, i.e. $\sqrt{U_i\Gamma_I}$, must be the same on both dots.
However this is not a serious constraint as we are much more interested in physics occurring on a scale
associated with spin fluctuations which are not so constrained.

In the bulk of the paper, we focus this approach on studying double dots (i.e. $\alpha=1,2$ in the
above model).
We identify a number of interesting phenomena.  At the particle-hole symmetric point of the double dot system
we find that the ground state
of the system is a singlet {\it and} is characterized by a Kondo-like resonance. 
This implies that the RKKY interaction between the dots acts in an unusual fashion.  In the 
standard picture, one 
expects either that an antiferromagnetic RKKY interaction binds the two electrons into a singlet but with any Kondo effect
absent or that a ferromagnetic RKKY interaction binds the electrons into a triplet
which is subsequently reduced upon Kondo screening to a doublet.  But here we find neither scenario.
Instead we find an effective antiferromagnetic RKKY interaction that cooperates, not competes, at an equivalent
energy scale with Kondo singlet formation.
We term this an RKKY Kondo effect.  This RKKY effect is clearly much different than the perturbative
ferromagnetic interaction \cite{kittel} between dots appropriate for a 
high temperature analysis of the dot
system.\cite{utsumi}
Using the Bethe ansatz, we are able to describe
quantitatively
the associated Abrikosov-Suhl resonance together with how it evolves under applications of finite
gate voltage (thus destroying particle-hole symmetry), finite magnetic field, and finite temperature.

We also identify a phenomena that we believe is underappreciated.  At values of the gate
voltage where the dot system is transitioning from the mixed valence regime to the empty orbital
regime, the number of electrons displaced, $n_{\rm dis}$, in connecting the leads to the dots is negative.
$n_{\rm dis}$ is a sum of both the localized electrons on the dots and changes in the electron density
in the leads.  Our analysis allows us to argue that this second contribution to $n_{\rm dis}$ is 
finite (and negative).  This phenomena is associated in general with the presence of interference
and for example has also been predicted to occur in dots exhibiting Fano resonances.\cite{fano}

A detailed outline of the paper is as follows.  It is divided in two overall parts.  The first
part, comprising Sections II through IV, is technical in nature.  In Section II we both introduce the  
model of N-dots in parallel and demonstrate that it admits an exact solution via the Bethe ansatz.
To do so we construct both the one and two particle wavefunctions explicitly.  This allows the extraction
of the S-matrix and thus the ability to construct {\it implicitly}
the N-particle wavefunctions through the algebraic Bethe ansatz.  We also sketch briefly the {\it explicit}
construction of the N-particle wavefunctions.  This is necessary to ensure the posited Bethe ansatz
wavefunctions are indeed eigenfunctions of the dot Hamiltonian.  Finally in Section II we provide a general
analysis of the Bethe ansatz equations in the continuum limit both at zero and finite temperature.  

In Section III we sketch how the transport properties of the double dot can be accessed via the Bethe ansatz
equations.  We do so on the same basis that the magnetoconductance was computed in the Kondo model.\cite{andrei}
There the scattering phase of an electron off the spin impurity was computed by analyzing the $1/L$ (where $L$ is the
system size) corrections of the Bethe ansatz equations.  Here the analysis is more involved as Anderson-type
models have non-trivial excitations involving both charge and spin degrees of freedom (unlike the Kondo model
where the charge sector decouples).  We, however, are still able to relate the scattering phase to an impurity
density of states.  As this latter quantity is readily extracted from the Bethe ansatz equations, we are
in position to characterize the transport through the dot.  In Section III we sketch how this proceeds both
at zero and finite temperature.

Up to this point in the article, we
have made no attempt to provide specific solutions to the integral equations arising out of the analysis
of the Bethe ansatz quantization condition.  In
Section IV of the paper this changes.   We both specialize to the case of two dots and we outline how the
equations describing transport through two-dot systems can be solved analytically in the two particular limits
of large and small bare level separation, i.e. $\delep \gg V_{1,2}^2$ and $\delep \ll V_{1,2}^2$.
This ends
the technical portion of the paper.

In the second part of the paper, Sections V through VII, 
we discuss the various features of transport arising in a double parallel dot system.
These sections can be read (mostly) independently from the Bethe ansatz analysis in the paper's first
part.  In Section V we study zero temperature transport both in and out of an applied Zeeman field.
In zero field we study the conductance through the double dots as a function of gate voltage.  We identify
three regions of gate voltage: i) near the particle-hole symmetric point where two electrons sit on
the dots and where what we term an RKKY Kondo
effect is present; ii) a region in gate voltage where one electron sits on the dots and a standard Kondo
effect takes place; and iii) a region in gate voltage marking the transition between a mixed valence 
and empty orbital regime where marked interference effects are present.  We then study how these various
phenomena are affected by the introduction of a Zeeman field.
In Section VI we study the finite temperature linear response conductance at the particle-hole symmetric point (i.e.
how the introduction of a temperature destroys the RKKY Kondo effect).  In Section VII we end the paper
considering two items.  
We first examine how the Friedel sum rule operates in the region where strong interference effects are present.
And second we consider how breaking the integrability constraints, i.e. Eqns. (\ref{eIii}) and (\ref{eIiii}),
affects the physics we have discussed in the manuscript.  On this last point we conclude that the
effects are in general perturbative, i.e. small violations of the constraints leads only to small changes
in the physics.

\section{Exact Solvability of N-Dots in Parallel}

\subsection{Integrability of Model: Bethe Ansatz Equations}

To begin to analyze the model in Eqn. (1.1) we first map the problem to an Anderson model
involving a single effective lead.  To do so we need to assume the ratio
of left/right lead couplings are equal, i.e. $V_{1\alpha}/V_{2\alpha}=V_{1\alpha'}/V_{2\alpha'}$.
Writing $c_{e/o} = (V_{1/2\alpha}c_1 \pm V_{2/1\alpha}c_2)/\sqrt{2\Gamma_\alpha}$, with 
$\Gamma_\alpha = (V_{1\alpha}^2+V_{2\alpha}^2)/2$, the Hamiltonian factorizes
into an even and an odd sector:
\begin{eqnarray}\label{eIIii}
{\cal H}_e &=& -i\sum_{l\sigma}\int^\infty_{-\infty} dx ~c^\dagger_{e\sigma}\partial_x c_{e\sigma}
+ \sum_{\sigma\alpha}\sqrt{2\Gamma_{\alpha}}(c^\dagger_{e\sigma\alpha}d_{\sigma\alpha} + {\rm h.c}) \cr\cr
&& + \sum_{\sigma\alpha}\epsilon_{d\alpha}n_{\sigma\alpha} + 
\sum_{\alpha\alpha'} U_{\alpha\alpha'} n_{\ua\alpha}n_{\da\alpha};\cr
{\cal H}_o &=& -i\sum_{l\sigma}\int^\infty_{-\infty} dx ~c^\dagger_{o\sigma}\partial_x c_{o\sigma}.
\end{eqnarray}
Only the even Hamiltonian, coupling directly to the dot degrees of freedom, is non-trivial.

To determine under what conditions ${\cal H}_e$ is exactly solvable by Bethe ansatz we compute both the
one and two electron eigenstates.  The one particle wave function takes the form,
\begin{eqnarray}\label{eIIiii}
|\psi_\sigma\rangle = \bigg[ \int^\infty_{-\infty}dx \{ g_\sigma (x) c^\dagger_\sigma(x) \} 
+ e_{\alpha\sigma} d^\dagger_{\alpha\sigma}\bigg]|0\rangle .
\end{eqnarray}
We solve the Schr\"odinger equation, ${\cal H}_e |\psi\rangle = q|\psi\rangle$, and find
$g(x)$ is of the form,
\begin{equation}\label{eIIiv}
g_\sigma (x) = \theta (x) e^{iqx + i\frac{\delta(q)}{2}} + \theta (-x) e^{iqx - i\frac{\delta(q)}{2}},
\end{equation}
with $\delta (q)$ the scattering phase of the electron off the dot array
equaling,
\begin{equation}\label{eIIv}
\delta (q) = -2\tan^{-1}(\sum_\alpha \frac{\Gamma_\alpha}{q - \epsilon_{d\alpha}} ).
\end{equation}
The total scattering phase is a sum of the {\it arguments} of $\tan^{-1}$ of
scattering phases off the individual dots in the system.

The two particle eigenfunction in the $S_z=0$ sector takes the form 
\begin{eqnarray}\label{eIIvi}
|\psi\rangle &=& \bigg[ \int^{\infty}_{-\infty} dx_1dx_2 g(x_1,x_2)c^\dagger_\uparrow (x_1) 
c^\dagger_\downarrow (x_2)
+ \sum_{\alpha}\int^\infty_{-\infty} dx  \cr\cr
&& \hskip -.45in 
\big[e_\alpha(x) (c^\dagger_\uparrow (x) d^\dagger_{\alpha\da} - c^\dagger_\downarrow (x)d^\dagger_{\ua\alpha})\big]
\!\!+\!\! \sum_{\alpha\alpha'} f_{\alpha\alpha'}d^\dagger_{\ua\alpha} d^\dagger_{\da\alpha'}\bigg]|0\rangle .
\end{eqnarray}
Again solving the Schr\"odinger equation ${\cal H}_e|\psi\rangle = (q+p)|\psi\rangle$ gives $g(x_1,x_2)$ to be
\begin{eqnarray}\label{eIIvii}
g(x_1,x_2) &=& g_q (x_1)g_p(x_2)\phi(x_1-x_2)\cr\cr 
&&\hskip .3in + g_p(x_1)g_q(x_2)\phi(x_2-x_1).
\end{eqnarray}
Here $g_{q/p}(x)$ and $e_{q/p}$ are the coefficients appearing in the one particle wavefunction corresponding to
energies $q/p$.  $\phi (x)$ governs the scattering when two electrons are interchanged.  It takes the
form $\phi(x) = 1 + i\gamma (q,p){\rm sign} (x)$.  We find that
$\gamma (q,p)$ must satisfy for arbitrary $\alpha$,
\begin{eqnarray}\label{eIIviii}
\gamma (q,p) &=& \frac{1}{2}\frac{1}{q - p} \sum_{\alpha'} 
\frac{p_{\alpha}q_{\alpha'}+p_{\alpha'}q_{\alpha}}{p_{\alpha'} q_{\alpha'}} \cr\cr
&& \times \frac{2\Gamma_{\alpha'}U_{\alpha\alpha'}}{\epsilon_{d\alpha} +
\epsilon_{d\alpha'}+U_{\alpha\alpha'} - q - p},
\end{eqnarray}
where $p_{\alpha}/q_{\alpha} = p-\epsilon_{d\alpha}/q-\epsilon_{d\alpha}$.
For consistency, $\gamma (p,q)$ must be independent of a particular vale of $\alpha$.  This
only happens if one of two sets of conditions holds:
\vskip 10pt
\noindent case i:
\begin{eqnarray}\label{eIIix}
U_{\alpha\alpha '} &=& \delta_{\alpha \alpha'}U_{\alpha};~~U_{\alpha} \Gamma_\alpha = 
U_{\alpha'}\Gamma_{\alpha'};\cr\cr
&& \hskip -.14in U_{\alpha} + 2\epsilon_{d\alpha} = U_{\alpha'} + 2\epsilon_{d\alpha'};
\end{eqnarray}
\vskip 10pt
\noindent case ii:
\begin{eqnarray}\label{eIIx}
U_{\alpha\alpha'} &=& U;~~\Gamma_\alpha = \Gamma_{\alpha'};~~\epsilon_{d\alpha} = \epsilon_{d\alpha'} = \epsilon_d.
\end{eqnarray}
Case i) would be appropriate to describing well separated single level dots while case ii) would
be appropriate to describing a single dot with degenerate levels.

Exact solvability is predicated on how $\gamma (q,p)$ determines the scattering matrix of the two electrons.
The scattering matrix has the general spin (SU(2)) invariant form
\begin{eqnarray}\label{eIIxi}
S^{\sigma_1'\sigma_2'}_{\sigma_1\sigma_2} = b(p,q) I^{\sigma_1'\sigma_2'}_{\sigma_1\sigma_2}
+ c(p,q) P^{\sigma_1'\sigma_2'}_{\sigma_1\sigma_2},
\end{eqnarray}
where here $\{\sigma_1,\sigma_2$\}/\{$\sigma_{1'},\sigma_{2'}\}$ represent the incoming and outgoing spins.
Here $I^{\sigma_1'\sigma_2'}_{\sigma_1\sigma_2} = \delta^{\sigma_1'}_{\sigma_1}\delta^{\sigma_2'}_{\sigma_2}$
is the identity matrix while 
$P^{\sigma_1'\sigma_2'}_{\sigma_1\sigma_2} = \delta^{\sigma_2'}_{\sigma_1}\delta^{\sigma_1'}_{\sigma_2}$ is
the permutation matrix.  The coefficients, $b(p,q)$ and $c(p,q)$, are determined by $\gamma (p,q)$ from
the two relations
\begin{eqnarray}\label{eIIxii}
b(p,q)-c(p,q) &=& \frac{\phi(x_1-x_2)}{\phi(x_2-x_1)} = \frac{1+i\gamma (p,q)}{1-i\gamma(p,q)}\cr\cr
b(p,q)+c(p,q) &=& 1.
\end{eqnarray}
The first relation results from the two particle eigenfunction in the $S_z=0$ sector solved above 
while the second relation arises from the corresponding eigenfunction in the $S_z = \pm 1$ sector
where interactions between spins are trivial.

\begin{figure}[ht]
\begin{center}
\epsfxsize=0.4\textwidth
\epsfbox{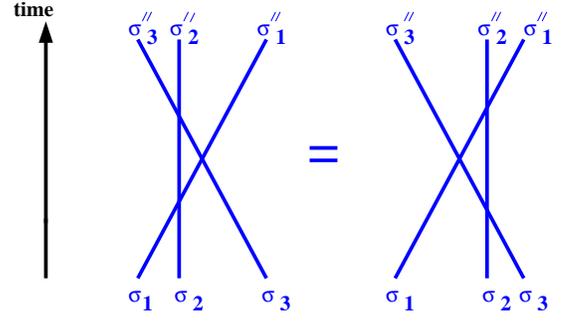}
\end{center}
\caption{A graphical representation of the Yang-Baxter equation.  Three particles $\sigma_1,\sigma_2$,
and $\sigma_3$ scatter into $\sigma_1'',\sigma_2'',$ and $\sigma_3''$ in two different ways.
Under the Yang-Baxter relation, the two scattering processes are equivalent.}
\end{figure}

In order for the Hamiltonian ${\cal H}_e$ to be integrable, the above S-matrix must satisfy the
Yang-Baxter relation.  The Yang-Baxter relation governs the scattering of three electrons,
$\sigma_1,\sigma_2$, and $\sigma_3$.  It enforces the equivalency of different scattering orders.
$\sigma_1$ may first scatter off $\sigma_2$ and then $\sigma_3$ followed by a scattering of $\sigma_2$
and $\sigma_3$ or $\sigma_2$ and $\sigma_3$ may scatter first followed by successive scattering
of $\sigma_1$ off $\sigma_2$ and $\sigma_3$.  This is graphically encoded in Figure 2.  Algebraically
it is given by
\begin{eqnarray}\label{eIIxiii}
&& S^{\sigma_1'\sigma_2'}_{\sigma_1\sigma_2}(p_1,p_2)S^{\sigma_1''\sigma_3'}_{\sigma_1'\sigma_3}(p_1,p_3)
S^{\sigma_2''\sigma_3''}_{\sigma_2'\sigma_3'}(p_2,p_3) =\cr\cr
&& \hskip .15in  S^{\sigma_2'\sigma_3'}_{\sigma_2\sigma_3}(p_2,p_3)S^{\sigma_1'\sigma_3''}_{\sigma_1\sigma_3'}(p_1,p_3)
S^{\sigma_1''\sigma_2''}_{\sigma_1'\sigma_2'}(p_1,p_2).
\end{eqnarray}
If this relation holds, the scattering among three particles is independent
of the order in which the particles scatter among one another.  For an SU(2) invariant system, the
Yang-Baxter relationship reduces to the simple condition,\cite{wie}
\begin{eqnarray}\label{eIIxiv}
b(p,q)/c(p,q) = g(p) - g(q),
\end{eqnarray}
where $g(p)$ is some function.    In case i), $g(p)$ is given by
\begin{equation}\label{eIIxv}
g(p) = \frac{(p - \epsilon_{d\alpha} - U_{\alpha\alpha}/2)^2}{2\Gamma_\alpha U_{\alpha\alpha}} ;
\end{equation}
while in case ii), $g(p)$ equals
\begin{equation}\label{eIIxvi}
g(p) = \frac{(p - \epsilon_d - U/2)^2}{2U \sum_\alpha \Gamma_\alpha}.
\end{equation}
In the second case, $g(p)$ is equivalent to a single dot with level broadening $\Gamma = \sum_\alpha \Gamma_\alpha$.

Having determine under 
what conditions the dot-lead Hamiltonian is exactly solvable, we are now in a position to
construct 
$N$-particle eigenstates in a controlled fashion.  An eigenfunction with spin $2S_z = N-2M$, 
is characterized
by a sea of N electrons each carrying momenta $\{ q_i\}^N_{i=1}$ and so total energy $E=\sum_i q_i$.  
In a periodic system of length L, integrability allows us to write down in a compact form the 
$k_i$-quantization conditions (the Bethe ansatz equations):
\begin{eqnarray}\label{eIIxvii}
e^{iq_jL+i\delta(q_j)} &=& \prod^M_{\alpha = 1}\frac{g(q_j)-\lambda_\alpha+i/2}{g(q_j)-\lambda_\alpha-i/2};\cr\cr
\prod^N_{j=1}\frac{\lambda_\alpha-g(q_j)+i/2}{\lambda_\alpha-g(q_j)-i/2} 
&=& -\prod^M_{\beta=1}\frac{\lambda_\alpha-\lambda_\beta+i}{\lambda_\alpha-\lambda_\beta-i}.
\end{eqnarray}
These equations are identical to those for the ordinary Anderson model\cite{wie} but for the form of $\delta (q)$
(here given in Eqn. (\ref{eIIv})).  The M-$\lambda_\alpha$ appearing in the above equations are indicative of the
spin degrees of freedom.  

The Bethe ansatz equations are derived in the same fashion (using the algebraic Bethe ansatz)
as described in Ref. (\onlinecite{wie}).  
However it is important to also explicitly construct the N-particle wavefunctions.
There is an {\it a priori} possibility that even though the integrability conditions (i.e. Eqn. (\ref{eIIxiv}))
are satisfied, the N-particle wavefunctions posited by the Bethe ansatz are not actual eigenfunctions
of the Hamiltonian.  This possibility has been seen previously in Hubbard models with more
than two species of electrons.\cite{halchoy}
In such a case what spoils integrability is the possibility that more than two electrons sit on the same
site.  A similar possibility thus exists here where more than two electrons can sit on the dots.  

\newcommand{\dd}{\dagger}
\newcommand{\cu}{c^\dagger_\uparrow}
\newcommand{\cd}{c^\dagger_\downarrow}

For simplicity, we only consider the case where two dots are involved.  The N-particle wavefunctions 
with M $\downarrow$ electrons (assuming $N>M>2$) in this case take the form
\begin{widetext}
\begin{eqnarray}\label{eIIxviia}
|\psi_{N,M}\rangle &=& \int dx_1 \cdots dx_N g(x_1,\ldots,x_N)\cd(x_1)\cdots\cd(x_M)\cu(x_{M+1})\cdots\cu(x_N)\cr\cr
&+& \sum_{\alpha=1,2,\sigma=\ua,\da}\int dx_1 \cdots dx_{M-s_\da(\sigma)}dx_{M+1}\cdots dx_{N-s_\da(\sigma)} 
e_{\sigma\alpha}(x_1,\ldots,x_{M-s_\da(\sigma)},x_{M+1},\ldots,x_{N-s_\ua(\sigma)})\cr
&&\hskip 2.in \times\cd(x_1)\cdots\cd(x_{M-s_\da(\sigma)})\cu(x_{M+1})\cdots\cu(x_{N-s_\ua(\sigma)})
{d^\dd}_{\sigma\alpha}\cr\cr
&+& \sum_{\sigma\sigma'\alpha\alpha'}\int dx_1 \cdots dx_{M-s_\da(\sigma)-s_\da(\sigma')} dx_{M+1}\cdots 
dx_{N-s_\da(\sigma)-s_\da(\sigma')}\cr\cr
&& \hskip 1in \times 
f_{\sigma\sigma'\alpha\alpha'}(x_1,\ldots,x_{M-s_\da(\sigma)-s_\da(\sigma')},x_{M+1},\ldots,x_{N-s_\ua(\sigma)-s_\ua(\sigma')})\cr\cr
&&\hskip 1in \times\cd(x_1)\cdots\cd(x_{M-s_\da(\sigma)-s_\da(\sigma')})\cu(x_{M+1})\cdots\cu(x_{N-s_\ua(\sigma)-s_\ua(\sigma')})
{d^\dd}_{\sigma\alpha} {d^\dd}_{\sigma'\alpha'}\cr\cr
&+&\sum_{\sigma\sigma'\sigma''\alpha\alpha'\alpha''}\int dx_1 \cdots dx_{M-s_\da(\sigma)-s_\da(\sigma')-s_\da(\sigma'')} 
dx_{M+1}\cdots dx_{N-s_\ua(\sigma)-s_\ua(\sigma')-s_\ua(\sigma'')}\cr\cr
&& \hskip .2in r_{\sigma\sigma'\sigma''\alpha\alpha'\alpha''}(x_1,\ldots,x_{M-s_\da(\sigma)-s_\da(\sigma')-s_\da(\sigma'')}
,x_{M+1},\ldots,x_{N-s_\ua(\sigma)-s_\ua(\sigma')-s_\ua(\sigma'')})\cr\cr
&&\hskip .2in \times\cd(x_1)\cdots\cd(x_{M-s_\da(\sigma)-s_\da(\sigma')-s_\da(\sigma'')})\cu(x_{M+1})\cdots\cu(x_{N-s_\ua(\sigma)-s_\ua(\sigma')-s_\ua(\sigma'')})
{d^\dd}_{\sigma\alpha} {d^\dd}_{\sigma'\alpha'}{d^\dd}_{\sigma''\alpha''}\cr\cr
&+&\int dx_1 \cdots dx_{M-2} dx_{M+1}\cdots dx_{N-2}\
h(x_1,\ldots,x_{M-2},x_{M+1},\ldots,x_{N-2})\cr
&&\hskip 2.in \times\cd(x_1)\cdots\cd(x_{M-2})\cu(x_{M+1})\cdots\cu(x_{N-2})
{d^\dd}_{\ua 1} {d^\dd}_{\da 1}{d^\dd}_{\ua 2} {d^\dd}_{\da 2},
\end{eqnarray}
\end{widetext}
where $s_\da(\da)=s_\ua(\ua)=1$ and $s_\da(\ua)=s_\ua(\da) = 0$.
The different terms of the wavefunction correspond to different numbers of electrons sitting on the dots.
We look for eigenstates with total energy $E=\sum^N_{i=1}q_i$ where the $q_i$ are single particle
energies.
Under the Bethe ansatz the coefficients, $g(x_i)$, $e_{\sigma\alpha}(x_i)$, $f_{\sigma\sigma'\alpha\alpha'}(x_i)$,
$r_{\sigma\sigma'\sigma''\alpha\alpha'\alpha''}(x_i)$, and $h(x_i)$ take the form
\begin{widetext}
\begin{eqnarray}\label{eIIxviib}
g(x_1,\ldots,x_N) &=& \sum_{\sigma\in S_N}\prod g_{q_{\sigma (i)}}(x_i) \cdot A_{N,M}(x_i|\sigma);\cr\cr
e_{\sigma\alpha}(x_1,\ldots,x_{M-s_\da(\sigma)},x_{M+1},\ldots,x_{N-s_\ua(\sigma)}) 
&=& \sum_{\sigma\in S_N}
\prod^{M-s_\da(\sigma)}_{i=1}g_{q_{\sigma (i)}}(x_i) 
\prod^{N-s_\ua(\sigma)}_{i=M+1}g_{q_{\sigma (i)}}(x_i) \cr\cr
&& \hskip -1.5in \times
\prod^{M}_{i=M-s_\da(\sigma)+1} e_{q_{\sigma(i)}}
\prod^{N}_{i=N-s_\ua(\sigma)+1} e_{q_{\sigma(i)}}
B_{\sigma\alpha N,M}(x_1,\ldots,x_{M-s_\da(\sigma)},x_{M+1},
\ldots,x_{N-s_\ua(\sigma)}|\sigma);\cr\cr
&& \hskip -2.5in f_{\sigma\sigma'\alpha\alpha'}(x_1,\ldots,x_{M-s_\da(\sigma)-s_\da(\sigma')},x_{M+1},
\ldots,x_{N-s_\ua(\sigma)-s_\ua(\sigma')}) 
= \cr\cr
&& \hskip -1.5in \times \sum_{\sigma\in S_N}
\prod^{M-s_\da(\sigma)-s_\da(\sigma')}_{i=1}g_{q_{\sigma (i)}}(x_i) 
\prod^{N-s_\ua(\sigma)-s_\ua(\sigma')}_{i=M+1}g_{q_{\sigma (i)}}(x_i) \cr\cr
&& \hskip -1.5in \times
\prod^{M}_{i=M-s_\da(\sigma)-s_\da(\sigma')+1} e_{q_{\sigma(i)}}
\prod^{N}_{i=N-s_\ua(\sigma)-s_\ua(\sigma')+1} e_{q_{\sigma(i)}}
\cr\cr
&& \hskip -1.5in \times C_{\sigma\sigma'\alpha\alpha' N,M}
(x_1,\ldots,x_{M-s_\da(\sigma)-s_\da(\sigma')},x_{M+1},\ldots,x_{N-s_\ua(\sigma)-s_\ua(\sigma')}|\sigma);\cr\cr
&& \hskip -2.5in 
r_{\sigma\sigma'\sigma''\alpha\alpha'\alpha''}(x_1,\ldots,x_{M-s_\da(\sigma)-s_\da(\sigma')-s_\da(\sigma'')},x_{M+1},
\ldots,x_{N-s_\ua(\sigma)-s_\ua(\sigma')-s_\ua(\sigma'')}) = \cr\cr
&& \hskip -1.5in \times \sum_{\sigma\in S_N}e_{q_{\sigma(i)}}
\prod^{M-s_\da(\sigma)-s_\da(\sigma')-s_\da(\sigma'')}_{i=1}g_{q_{\sigma (i)}}(x_i) 
\prod^{N-s_\ua(\sigma)-s_\ua(\sigma')-s_\ua(\sigma'')}_{i=M+1}g_{q_{\sigma (i)}}(x_i) \cr\cr
&& \hskip -1.5in \times
\prod^{M}_{i=M-s_\da(\sigma)-s_\da(\sigma')-s_\da(\sigma'')+1} e_{q_{\sigma(i)}}
\prod^{N}_{i=N-s_\ua(\sigma)-s_\ua(\sigma')-s_\ua(\sigma'')+1} e_{q_{\sigma(i)}}\cr\cr
&& \hskip -1.5in \times D_{\sigma\sigma'\sigma''\alpha\alpha'\alpha'' N,M}
(x_1,\ldots,x_{M-s_\da(\sigma)-s_\da(\sigma')-s_\da(\sigma'')},x_{M+1},\ldots,x_{N-s_\ua(\sigma)-s_\ua(\sigma')
-s_\ua(\sigma'')}|\sigma);\cr\cr
&& \hskip -2.5in h(x_1,\ldots,x_{M-2},x_{M+1},\ldots,x_{N-2}) = 
\sum_{\sigma\in S_N}e_{q_{\sigma(i)}}
\prod^{M-2}_{i=1}g_{q_{\sigma (i)}}(x_i) 
\prod^{N-2}_{i=M+1}g_{q_{\sigma (i)}}(x_i) \cr\cr
&& \hskip .2in \times 
E_{N,M}(x_1,\ldots,x_{M-2},x_{M+1},\ldots,x_{N-2}|\sigma).
\end{eqnarray}
\end{widetext}
Here $\sum_{\sigma \in S_N}$ 
is a sum over the permutations of $N$.  $g_q(x)$ is the electron portion of the single
particle eigenstate defined in (\ref{eIIiv}) with energy $q$.  Similarly $e_q$ is the dot 
portion of this same wavefunction.  
The coefficients $A_{N,M}$ through $E_{N,M}$ depend upon both $\sigma$
and the relative ordering of the $x_i$ (for example, $A_{N,M}$ only changes value as the $x_i$ cross
one another).  Their forms are analogous to those of the Hubbard model detailed in Ref. (\onlinecite{hubbard}). 

By direct substitution we check that this posited form of the wavefunction does indeed correspond to an eigenfunction
of the Hamiltonian.  While straightforward, the associated algebra is tedious and so is suppressed.

While we have only consider two dots in parallel, we can easily generalize to N-dots.  In order for
N-dots to be integrable we again require that $V_{1\alpha}/V_{2\alpha} = V_{1\alpha'}/V_{2\alpha'}$ for all
$\alpha$ and $\alpha'$ and that constraints analogous to those in Eqns. (\ref{eIIviii}) and (\ref{eIIix}) are 
satisfied.  This type of generalization is 
akin to the multi-lead generalizations of the single level Anderson
model found in Refs. \onlinecite{aus} and \onlinecite{simon1}.

\subsection{Analysis of Bethe Ansatz Equations}

\subsubsection{Description at $T=0$}

N-particle states with spin projection, $2S_z = N-2M$,
of the dot-lead system in general consists of a multitude of different types
of solutions for the $q$'s appearing in the Bethe ansatz quantization conditions of Eqn. (\ref{eIIxvi}).  
But at $T=0$ only two solutions are relevant for the ground state.  The ground state
at $T=0$ is composed of 1) N-2M real values of q and 2) 2M complex values of $q$ interlinked to $M$ real
values of $\lambda$ according to the rule
\begin{equation}\label{eIIxviii}
q_\pm(\lambda) = x(\lambda ) \pm i y(\lambda); ~~~ g(q_\pm(\lambda )) = \lambda \pm i/2.
\end{equation}
In the continuum limit, densities $\rho (q)$ and $\sigma (\lambda)$ describing the distributions of $q$ and $\lambda$ 
respectively may be derived\cite{wie}
\begin{eqnarray}\label{eIIxix}
\rho (q) &=& \frac{1}{2\pi} + \frac{\Delta (q)}{L} + 
g'(q) \il a_1(g(q)-\la) \sig (\la); \cr\cr
\sig (\la ) &=& - \frac{x'(\la)}{\pi} + \frac{\tilde{\Delta}(\la)}{L}
- \ilp a_2(\la '-\la)\sig (\la ')\cr\cr
&& - \int^B_{-D}dq a_1(\la - g(q))\rho (q),
\end{eqnarray}
where
\begin{eqnarray}\label{eIIxx}
\Delta (q) &=& \frac{1}{2\pi} \partial_q \delta (q);\cr\cr
\tilde{\Delta} (\la ) &=& -\frac{1}{\pi} \del_\la 
{\rm Re}\delta (x(\la)+iy(\la));\cr\cr
a_n(x) &=& \frac{1}{2\pi}
\partial_x \theta_n (x) =  \frac{2n}{\pi} \frac{1}{(n^2 + 4x^2)}.
\end{eqnarray}
In the equations for the charge, $\rho(q)$, and spin, $\sigma(\lambda)$, 
distributions appear the Fermi surfaces, $Q$ and $B$, and ``band'' bottoms
$\tilde Q$ and $-D$.  These mark out the range of $q$ and $\lambda$ over which excitations appear in the ground
state. $-D$ is the lower band edge of the charge excitations.  As each $\lambda$ has two associated complex $q$'s,
we expect $q_{q_+}(\tilde Q) + q_{q_-}(\tilde Q) = 2x(\tilde Q) = -2D$, thus determining $\tilde{Q}$.\cite{wie}  The Fermi
surfaces $Q$ and $B$ are fixed by insisting that the overall spin and particle number,
\begin{eqnarray}\label{eIIxxi}
N-2M &=& L\int^B_{-D} dq \rho (q);\cr\cr
M &=& L\int^{\tilde Q}_{Q} d\lambda \sigma (\lambda ) ,
\end{eqnarray}
are reproduced.

In understanding transport properties we will want to construct electronic-like excitations (even if
zero-energy) about this ground state.  These excitations are in turn related to specific solutions,
$q$ and $\lambda$, of the Bethe ansatz equations.  If for example we add an electron to the system,
i.e. take $N \rightarrow N+1$, to effect this change, we must add a real $q$-state as well as a $\lambda$-hole.
Thus we must parameterize the energies, $\epsilon_q(q)$ and $\epsilon (\lambda)$ of the these excitations.  
Equations governing these energies
are easily derivable (see Ref. (\onlinecite{long}))
and are given by
\begin{eqnarray}\label{eIIxxii}
\epsilon_q(q) &=& q - \frac{H}{2} - \int^{\tilde{Q}}_Q d\lambda \epsilon_\lambda (\lambda ) a_1(\lambda - g(q));\cr\cr
\epsilon_\lambda (\lambda ) &=&  2 x(\lambda ) - \int^{\tilde{Q}}_Q d\lambda' 
\epsilon_\lambda (\la ' )a_2(\la ' - \la)\cr\cr
&&+ \int^B_{-D} g'(q)\epsilon_q(q) a_1(g(q)-\la).
\end{eqnarray}
$\epsilon_q (q)$ is a monotonically increasing function that is equal to zero at the $q$-Fermi
surface, i.e. $\epsilon_q(B) = 0$.  For $q>B$, $\epsilon_q (q) > 0$ indicating it costs
energy, $\epsilon_q (q)$, to add a state above the $q-$Fermi surface while for $q<B$, $\epsilon_q(q) <0$, indicating
it costs energy to add a hole below the Fermi surface.  Similar consideration hold for $\epsilon_\lambda$
(although here $\epsilon (\lambda)$ is a monotonically decreasing function which passes through 0
at $\lambda = Q$).

\subsubsection{Description at Finite $T$}

The description of the system at finite temperature is considerably more involved.  At finite $T$, an infinite
hierarchy of excitations of the Bethe ansatz equations must be considered.  The two coupled integral equations
at $T=0$ are replaced by an a correspondingly infinite set.  These equations are closely related to those
analyzed in Ref. (\onlinecite{wie,long}).  However some important differences are present and so we give
the description of the system at finite temperature in some detail.

At finite temperature, the possible excitations of the system fall into three classes
\vskip .1in 
\noindent {\bf i) real q:} As we have already seen, such solutions to the Bethe ansatz equations appear in the zero temperature
ground state at finite H.
\vskip .1in
\noindent {\bf ii) n-spin complex with 2n associated complex q's:}  The n-spin complex groups together n different
$\lambda$'s according to the rule
$$
\lambda^{nj} = \lambda^n + i(\frac{n+1}{2} - j), ~~~ j=1,\cdots, n.
$$
Here $\lambda^n$, the centre of the complex, is a real number.  For each $\lambda^{nj}$, there are two associated
$q$'s:
\begin{eqnarray*}
g(q^{+nj}) &=& \lambda^n+ i(\frac{n}{2}+1-j);\cr\cr
g(q^{-nj}) &=& \lambda^n+ i(\frac{n}{2}-j).
\end{eqnarray*}
The simplest of this type of complex ($n=1$), as we have already seen, fills the $H=T=0$ ground
state.
\vskip .1in
\noindent {\bf iii) n-spin complex with no associated $q$'s:}  As in case ii), the spin complex groups together
n different $\lambda$'s according to 
$$
\lambda^{nj} = \lambda^n + i(\frac{n+1}{2} - j), ~~~ j=1,\cdots, n.
$$
However here there are no associated $q$'s.

These excitations are governed by the following set of density equations describing
their occupancies at finite T
(akin to Eqn. (\ref{eIIxvii}) for the simpler $T=0$ case)
\begin{eqnarray}\label{eIIxxiii}
\rho_p (q) + \rho_h (q) &=& \frac{1}{2\pi} + \frac{\Delta (q)}{L} \cr\cr
&& \hskip -1.1in + ~ g'(q) \sum^{\infty}_{n=1} \int^\infty_{-\infty} d\lambda a_n (g(q)-\la )
(\sigma_{pn}(\la ) + \sigma'_{pn} (\la ));\cr\cr
\sigma_{hn} (\la ) &=& -\frac{x_n'(\la )}{\pi}  +
\frac{\tilde{\Delta_n}(\la )}{L} \cr\cr  
&& \hskip -.5in -\int^\infty_{-\infty} a_n(\la - g(q))\rho_p(q)\cr\cr
&& \hskip -.5in -\sum^\infty_{m=1} 
\int^\infty_{-\infty}d\la ' A_{nm}(\la - \la')\sigma_{pm}(\la ');\cr\cr
\sigma'_{hn} (\la ) &=& \int^\infty_{-\infty} a_n(\la - g(q))\rho_p(q)\cr\cr
&& \hskip -.5in - \sum^\infty_{m=1} 
\int^\infty_{-\infty}d\la ' A_{nm}(\la - \la')\sigma'_{pm}(\la ');
\end{eqnarray}
where $x_n(\lambda )$ and $\tilde\Delta_n (\la)$ are defined by
\begin{eqnarray}\label{eIIxxiv}
x_n(\la ) &=&
\sqrt{2U\Gamma}{\rm Re}~(\la + i\frac{n}{2})^{1/2} + n(\frac{U}{2} +\ep_d)\cr\cr
\tilde{\Delta}_n(\la ) &=& -\frac{1}{\pi}\partial_\la \delta_n (\la )\cr\cr
&\equiv&
-\frac{1}{\pi} \partial_\la {\rm Re}~
\delta (-\sqrt{2U\Gamma}\,(\la +i\frac{n}{2})^{1/2}+U/2+\ep_d)\cr\cr
&& \hskip -.6in -\frac{1}{2\pi}\del_\lambda \sum^{n-1}_{k=1} 
\bigg\{\delta(-\sqrt{2U\Gamma}\,(\la+\frac{i}{2}(n-2k))^{1/2}+U/2+\ep_d)\cr\cr
&& \hskip -.5in + \delta(\sqrt{2U\Gamma}\,(\la+\frac{i}{2}(n-2k))^{1/2}+U/2+\ep_d))\bigg\}.
\end{eqnarray}
The kernels in the above density integral equations are
\begin{eqnarray}\label{eIIxxv}
a_n(\la ) &=& \frac{2n}{\pi} \frac{1}{(n^2+4\la^2)};\cr\cr
A_{nm}(\la ) &=& \delta_{nm}\delta(\la ) + a_{|n-m|}(\la) \cr\cr
&& \hskip -.5in + 2\sum^{{\rm min}(n,m)-1}_{k=1} a_{|n-m|+2k} (\la ) + a_{n+m}(\la ).
\end{eqnarray}
In the above,  $\rho_{p/h}(q)$ describe the particle/hole occupancy of the real $q-$excitations.  The total
density of states, $\rho (q)$, is equal to the sum of the two
$$
\rho (q) = \rho_p(q) + \rho_h (q).
$$
To compare with the $T=0$ density equations, we note that at $T=0$,
\begin{eqnarray*}
\rho_p (q) &=& \Theta(B-q)\rho (q);\cr\cr
\rho_h (q) &=& \Theta(q-B)\rho (q).
\end{eqnarray*}
$\sigma_{p/hn}(\la)$ similarly describe the finite temperature particle/hole occupancies of the 
n-spin complexes with associated complex $q$'s (here $\la$ refers to the string centre).  At $T=0$
we have for comparison
\begin{eqnarray}\label{eIIxxvi}
\sigma_{pn} (\lambda ) &=& 0, n>1;\cr\cr
\sigma_{p1} (\lambda ) &=& \sigma_1(\lambda )\Theta(\lambda-Q);\cr\cr
\sigma_{h1} (\lambda ) &=& \sigma_1(\lambda )\Theta(Q-\lambda).
\end{eqnarray}
Finally $\sigma'_{p/hn}(\lambda )$ describes the particle/hole densities of the n-spin complexes without
associated complex-q's.  At $T=0$, $\sigma'_{pn}(\lambda)$ are uniformly zero (none of these excitations
appear in the ground state).

The energies of creating the various types of excitations are indifferent to the form of the impurity scattering
phase (as at $T=0$ in Eqn. (\ref{eIIxx})).  Thus these energies are the same as for the single dot case
(see Ref. (\onlinecite{long})) and are given by
\begin{eqnarray}\label{eIIxxvii}
\ep_q(q) &=& q \cr\cr
&& \hskip -.7in +~ T\sum^{\infty}_{n=1} \int^\infty_{-\infty}
d\la \log (\frac{f(-\ep'_{\la n} (\la))}{f(-\ep_{\la n} (\la))}) a_n(\la - g(q));\cr\cr
\log (f(\ep_{\la n}(\la ))) &=& - \frac{2}{T} x_n(\la ) \cr\cr 
&& \hskip -.7in - \int^\infty_{-\infty}
dq g'(q) \log (f(-\ep_q (q))) a_n(g(q) -\la)\cr\cr
&& \hskip -.7in + \sum^{\infty}_{m=1} \int d\la' A_{nm}(\la -\la ')\log (f(-\ep_{\la m}(\la ')));\cr\cr
\log (f(\ep'_{\la n}(\la )))  &=& \cr\cr
&& \hskip -.7in - \int^\infty_{-\infty}
dq g'(q) \log (f(-\ep_q (q))) a_n(g(q)-\la)\cr\cr
&& \hskip -.9in + \sum^{\infty}_{m=1} \int d\la' A_{nm}(\la -\la ')\log (f(-\ep'_{\la m}(\la '))),
\end{eqnarray}
where $f(\ep ) = (1+\exp (\ep/T))^{-1}$ is the Fermi distribution.  These energies are related to
the various particle-hole densities via
\begin{eqnarray}\label{eIIxxviii}
\rho_{p/h}(q) &=& (\rho_p(q)+\rho_h(q))f(\pm \epsilon_q (q)) \equiv \rho (q) f(\pm \epsilon (q));\cr\cr
\sigma_{p/hn}(\la ) &=& (\sigma_{pn}(\la )+\sigma_{hn}(\la ))f(\pm \epsilon_{\la n} (\la )) \cr\cr
&\equiv & \sigma (\la) 
f(\pm \epsilon_{\la n} (\la ));\cr\cr
\sigma'_{p/hn}(\la ) &=& (\sigma'_{pn}(\la )+\sigma'_{hn}(\la ))f(\pm \epsilon'_{\la n} (\la )) \cr\cr
&\equiv & \sigma' (\la) f(\pm \epsilon'_{\la n} (\la ));
\end{eqnarray}
These equations are arrived at from minimizing the free energy of the system.  The solutions to 
Eqns. (\ref{eIIxxiii}) and (\ref{eIIxxvii}) represent the distribution of excitations in thermal equilibrium.

\section{Computing Transport Properties: Methodology}

Here we describe how to extract the scattering phases of electrons off the dot system from the
Bethe ansatz equations.  We do so both for the system at zero and finite temperature {\it in equilibrium}.
We believe that the results are obtained by this methodology are {\it exact} at $T=0$ and an {\it excellent
approximation} at finite $T$.  The nature of this finite $T$ 
approximation is discussed in detail in Ref. (\onlinecite{long}).
\subsection{General Considerations}

To compute transport properties we must translate the effects of the map between the $L/R$ and the even/odd
electrons.  Excitations in the even/odd picture scatter off the collection of dots with some pure scattering
phase, $\delta_{e/o}(E)$, where $E$ is the energy of the excitation.  This is true regardless of energy as there
are no inelastic processes to take into account.  The excitations are exact eigenstates of the system and not merely
asymptotic approximations.  The transmission/reflection ({\cal T/R}) amplitudes of an excitation in the $L/R$ basis 
are related
to the two scattering phases, $\delta_{e/o}$ via the following ansatz:
\begin{eqnarray}\label{eIIIi}
e^{i\delta_e(E)} &=& {\cal R}(E) + {\cal T}(E);\cr\cr
e^{i\delta_e(E)} &=& 1 = {\cal R}(E) - {\cal T}(E).
\end{eqnarray}
$\delta_o(E)$ is necessarily $0$ as the odd sector is uncoupled from the dots.

The computation of the $\delta_e(E)$ for a given excitation is described in detail in Ref. (\onlinecite{long}).
It is based on a technique used to compute the magnetoresistance in the Kondo model.\cite{andrei}  The scattering
phase is identified with the portion of an excitation's momentum scaling as $1/L$:
\begin{equation}\label{eIIIii}
p = p^{\rm bulk} + p^{\rm imp}/L = p^{\rm bulk} + \delta_e(E)/L.
\end{equation}
For the purposes of computing transport properties, we are interested in the scattering of
electronic excitations.  In the Bethe ansatz solution of this model, all excitations see a spin-charge
separation.  In order to construct an electronic excitation, we must glue together a charge excitation with
a spin excitation.  How this is done and potential (finite energy/temperature) pitfalls are again discussed in detail
in Ref. (\onlinecite{long}).  A spin $\ua$ electron can be treated as a composite of a right-moving real $q$-excitation
and a left moving $\lambda$-hole (here by a $\lambda$-hole we mean a hole in a 1-spin complex with associated
complex $q's$).  Thus
\begin{equation}\label{eIIIiii}
\delta_{{\rm electron}\ua} = p_q^{\rm imp}(q) + p_\la^{\rm imp} (\lambda ).
\end{equation}
The energy of this composite particle is
\begin{equation}\label{eIIIiv}
\ep_{\rm el} = \ep_q (q) - \ep_\la(\la ).
\end{equation}
The scattering of spin $\da$ electrons can be accessed through a particle-hole transformation.

\subsection{Zero Temperature}

At $T=0$, we are interested in zero energy excitations at the Fermi surface.  The conductance through
the dot is given by:
\begin{eqnarray}\label{eIIIv}
G &=& \frac{e^2}{h} \big(|T_\ua(E=0)|^2 + |T_\da(E=0)|^2\big) \cr
&& \hskip -.35in = \frac{e^2}{h} \big(\sin^2(\frac{\delta_{e\ua}(E=0)}{2}) + \sin^2(\frac{\delta_{e\da}(E=0)}{2})\big).
\end{eqnarray}
The zero energy composite excitation is constructed by choosing the $q-,\la$-excitations at their
Fermi surfaces $q=B/\la=Q$.  (At zero temperature $\ep_q(q=B)=\ep_\la(\la=Q)=0$.)

A simplification in the computation of scattering phases arises from a relationship between the impurity
momenta and the impurity densities.  The impurity density is simply the portion of the density scaling
as $1/L$, that is, the piece sensitive to the presence of the dots:
\begin{eqnarray}\label{eIIIvi}
\rho (q) &=& \rho_{\rm bulk}(q) + \rho^{\rm imp} (q);\cr\cr
\sigma (\lambda ) &=& \sigma_{\rm bulk}(\lambda ) + \sigma^{\rm imp} (\lambda ).
\end{eqnarray}
Thus $\rho^{\rm imp}(q)$ and $\sigma^{\rm imp}(\la)$ satisfy 
\begin{eqnarray}\label{eIIIvii}
\rho^{\rm imp} (q) &=& \Delta (q) + 
g'(q) \il a_1(g(q)-\la) \sig^{\rm imp} (\la); \cr\cr
\sig^{\rm imp} (\la ) &=& \tilde{\Delta}(\la)
- \ilp a_2(\la '-\la)\sig^{\rm imp} (\la ')\cr\cr
&& - \int^B_{-D}dq a_1(\la - g(q))\rho^{\rm imp} (q),
\end{eqnarray}
In Ref. (\onlinecite{long}), it is then shown that
\begin{equation}\label{eIIIviii}
\partial_q p^{\rm imp}_q(q) = 2\pi\rho^{\rm imp}(q);~~~~ 
\partial_\lambda p^{\rm imp}_{\la}(\lambda )  = -2\pi \sigma^{\rm imp}(\lambda ).
\end{equation}
Thus by computing the impurity densities we can access the scattering phases of the excitations.  
In particular we see
\begin{eqnarray}\label{eIIIix}
p^{\rm imp}_q (B) &=& 2\pi \int^B_{-D}dq \rho^{\rm imp}(q);\cr\cr
p^{\rm imp}_\la (Q) &=& 2\pi \int^{\tilde Q}_Q d\la \sigma^{\rm imp}(\la).
\end{eqnarray}
We point out that the integrated densities do not necessarily equal the number of electrons
sitting on the dots.  For example
\begin{eqnarray*}
\sum_i n_{\up i} + n_{\down i} \neq \int^B_{-D}dq \rho^{\rm imp}(q) + \int^{\tilde Q}_Q d\la \sigma^{\rm imp}(\la),
\end{eqnarray*}
as would be the case for a single dot.  Rather the r.h.s. of the above equation includes contributions
coming from $1/L$ changes in the electron density of the leads.  We will discuss this further in Sections V and VII.

\subsection{Finite Temperature}

At finite temperature, the calculation is less straightforward.  We must both take into account all the thermally excited
excitations and their dressing of the electron scattering phase.\cite{long}  
For a variety of technical reasons, this is only feasible at the symmetric point of the dots,
i.e. $U_{ii}/2+\epsilon_{di} = 0$ as explained in Ref. (\onlinecite{long}).  At finite T the electronic excitations
have a Fermi distribution and so the conductance, $G$, equals
\begin{equation}\label{eIIIx}
G = \frac{e^2}{h} \int dE \big(-\frac{\partial f}{\partial E}\big) \big(|T_\ua(E)|^2 + |T_\da(E)|^2\big) .
\end{equation}
To determine the scattering probabilities, $T_{\uparrow/\downarrow}(E) = \sin^2(\frac{\delta(E,T)}{2})$, 
we construct excitations above
the thermal ground state discussed in Section II.B.2 in analogy to what we did at $T=0$.  The electronic
excitations are built out of real $q$ and an $n=1$ spin-charge complex, $\la$ (in the language of Section II.B.2).
We have specifically
\begin{equation}\label{eIIIxi}
\delta_{el\up}(\ep=\ep_{el},T) = p^{\rm imp}_q(q) + p^{\rm imp}_{1\la}(\la).
\end{equation}
However here $p^{\rm imp}_q(q)$ and $p^{\rm imp}_1(\la)$ are dressed by the presence of the other
thermal excitations in the ground state.  The finite temperature form of $p^{\rm imp}(q)$ and
$p^{\rm imp}_1(\la)$ is:\cite{long}
\begin{eqnarray}\label{eIIIxii}
p^{\rm imp}_q (q) &=& \delta (q) + \sum^\infty_{n=1} \int^\infty_{-\infty}
\theta_n (g(q) - \la ) \cr\cr
&& \hskip 1.2in \times (\sigma_{pn}^{\rm imp}(\la ) + 
\sigma_{pn}^{'\rm imp} (\la ));\cr\cr
p_{1\la}^{\rm imp} (\la ) &=& 2\delta_1 (\la ) 
+\int^\infty_{-\infty} dq \rho_p^{\rm imp} (q)\theta_1 (\la - g(q))\cr\cr
&& + \sum^\infty_{m=1} \int d\la ' 
\Sigma_{1m} (\la - \la ')\sigma_{pm}^{\rm imp}(\la '),
\end{eqnarray}
where the kernels $\theta_n(\la )$ and $\sigma_{1m}(\la)$ are given by
\begin{eqnarray}\label{eIIIxiii}
\theta_n(\la) &=& 2\tan^{-1}(\frac{2}{n}\la) - \pi;\cr\cr
\Sigma_{nm}(\la ) &=& \theta_{|n-m|}(\la) \cr\cr
&& \hskip -.8in + 2\sum^{{\rm min}(n,m)-1}_{k=1}\theta_{|n-m|+2k} (\la ) + 
\theta_{n+m}(\la ).
\end{eqnarray}
Comparing the above with Eqn. (\ref{eIIxxiv}), we find the relationship,
\begin{eqnarray}\label{eIIIxiv}
\partial_q p^{\rm imp}_q(q,T) = 2\pi \rho^{\rm imp}(q,T);\cr\cr
\partial_\la p^{\rm imp}_{1\la}(\la,T) = -2\pi \sigma^{\rm imp}_1(q,T)
\end{eqnarray}
Hence at finite temperature 
\begin{eqnarray}\label{eIIIxv}
\delta_{el\up}(T) &=& 2\pi \int^q_{-D}dq \rho^{\rm imp}(q,T)\cr\cr
&& \hskip .5in + 2\pi \int^{\tilde Q}_\la d\la \sigma^{\rm imp}_1(\la,T),
\end{eqnarray}
a generalization of Eqns. (\ref{eIIIiii}) and (\ref{eIIIix}).

We need to determine how to choose $q$ and $\lambda$ at finite temperature.  If we are 
interested in constructing an excitation of energy, $\epsilon_{el}$, out of a $q-$particle
and a $\lambda$-hole we must insist that
\begin{eqnarray}\label{eIIIxvi}
\epsilon_{el} = \epsilon_q(q) - \epsilon (\lambda).
\end{eqnarray}
This would seem to provide a two parameter space $(q,\lambda)$ for characterizing excitations.
However as explained in Ref. (\onlinecite{long}), we fix $\lambda$ to some value $\lambda_0$ allowing
only $q$ to vary.  One justification for doing so lies in the nature of the physics at the
symmetric point.  At the symmetric point, Kondo physics will be present in some form governed by
a scale, $T_K$.  Of the two energies only $\epsilon (q)$ sees variations on this scale.
$\epsilon (\lambda )$ in contrast see variations on the much larger energy scale, $\sqrt{U_{ii}\Gamma_i}$
governing charge fluctuations.  Thus it is natural to only vary $q$.  To fix $\lambda_0$ we exploit
the fact that at the symmetric point of the dot system, we expect the scattering phase to similarly
symmetric about zero energy, i.e.
\begin{equation}\label{eIIIxvii}
\delta_{el} (\epsilon,T ) = \delta_{el}(-\epsilon,T ).
\end{equation}
We thus have 
\begin{eqnarray}\label{eIIIxviii}
\delta_{el\up}(\ep_{el},T) &=& 2\pi \int^q_{-D}dq \rho^{\rm imp}(q,T)\cr\cr
&& \hskip .15in + 2\pi \int^{\tilde Q}_{\lambda_0} d\la \sigma^{\rm imp}_1(\la,T),
\end{eqnarray}
with $q$ chosen such than
\begin{eqnarray}\label{eIIIxviiia}
\epsilon_{el} &=& \epsilon (q) - \epsilon (\lambda_0).
\end{eqnarray}

To compute $\delta_{e\up}$ from Eqn. (\ref{eIIIxviii}), we need to compute $\rho^{\rm imp}(q)$ and
$\sigma^{\rm imp}_1(\la )$.  This can by done by solving Eqns. (\ref{eIIxxiii}) and (\ref{eIIxxvii}).  
However at the symmetric point
and at energies much smaller than $\sqrt{U_{ii}\Gamma_i}$, these equations can be simplified.  If we
recast the energies as
\begin{eqnarray}\label{eIIIxix}
\phi_n (\la ) &=& \frac{1}{T} \ep_n (\la - \frac{1}{\pi} \log (\frac{2 A}{T}));\cr\cr
\phi'_1 (g(q) ) &=& -\frac{1}{T} 
\ep (-g(q) + \frac{1}{\pi} \log (\frac{2 A}{T}));\cr\cr
\phi'_{n+1} (\la ) 
&=& \frac{1}{T} \ep'_n (-\la + \frac{1}{\pi} \log (\frac{2 A}{T}));\cr\cr
A &=& \frac{\sqrt{2U\Gamma}}{2\pi},
\end{eqnarray}
we can show that they satisfy the following set of integral equations
\begin{eqnarray}\label{eIIIxx}
\xi_n (\la ) \!\!&=&\!\! -\!\!\int^\infty_{-\infty} \!\!d\la' s(\la - \la ') 
\log (f(T\xi_{n-1}(\la )) f(T\xi_{n+1}(\la ))) \cr\cr
&& \hskip 1in - \delta_{n1} e^{\pi\la },
\end{eqnarray}
where $\xi_n = \phi_n$ or $\phi'_n$ and $s(\la ) = \cosh^{-1} (\pi \la )/2$.

The impurity density equations take the initial form given in Eqn. (\ref{eIIxxiii}).  If we
focus upon energies $\ll \sqrt{U_{ii}\Gamma_i}$, we can safely assume
\begin{equation}\label{eIIIxxi}
\sigma^{\rm imp}_{h1} (\lambda ) = 0 ; ~~~ \sigma^{\rm imp}_{m} = 0; ~~~ m>1.
\end{equation}
The impurity densities then reduce to 
\begin{eqnarray}\label{eIIIxxii}
\rho^{\rm imp}(q) &=& \Delta(q) \cr\cr
&& \hskip -.5in + g'(q) ~\sum^\infty_{n=1}\int^\infty_{-\infty}
d\la a_n(g(q)-\la)\sigma'^{\rm imp}_{pn}(\la );\cr\cr
&& \hskip -.5in + g'(q) ~\int^\infty_{-\infty} d\la a_1(g(q)-\la)\sigma^{\rm imp}_{p1}(\la );\cr\cr
\sigma^{\rm imp}_{h1} (\la ) &=& \tilde{\Delta}_1 (\la )
- \int^\infty_{-\infty} dq \rho^{\rm imp}_p(q) a_1(\la - g(q))\cr
&& - \int^\infty_{-\infty} d\la' a_2(\la - \la')\sigma^{\rm imp}_{p1}(\la' );\cr\cr
\sigma'^{\rm imp}_{hn} (\la ) &=&
 \int^\infty_{-\infty} dq \rho^{\rm imp}_{p}(q) a_n(\la - g(q))\cr\cr
&& \hskip -.5in -\int^\infty_{-\infty} d\la' \sum^\infty_{m=1} A_{nm}(\la - \la')
\sigma'^{\rm imp}_{pm}(\la' ).
\end{eqnarray}
Using the inverse
of the matrix $A_{nm}$,
\begin{equation}\label{eIIIxxiii}
A^{-1}_{nm} (\la ) = \delta_{nm}\delta(\la ) - s(\la ) 
(\delta_{nm+1}+\delta_{nm-1}),
\end{equation}
together with the relationship
\begin{eqnarray}\label{eIIIxxiv}
\delta_{n1} s(\la - \la'')
&=& \int^\infty_{-\infty}d\la' A^{-1}_{nm}(\la - \la ') a_m (\la' - \la''),\cr &&
\end{eqnarray}
we obtain
\begin{eqnarray}\label{eIIIxxv}
&&\hskip -1.5in \rho^{\rm imp}(q) 
= \Delta (q) + g'(q) \int^\infty_{-\infty} d\la s(\la - g(q)) 
\tilde{\Delta_1} (\la )\cr\cr
&&\hskip -.5in - g'(q)\int^\infty_{-\infty} d\la s(\la - g(q)) \sigma'^{\rm imp}_{h1}(\la );\cr\cr
\sigma^{\rm imp}_{p1}(\la) +\sigma^{\rm imp}_{h1}(\la) &=&
- \int^\infty_{-\infty}d\la' R(\la - \la')\tilde\Delta_1(\la')\cr\cr
&& \hskip -.85in +\tilde\Delta_1(\la') 
- \int^\infty_{-\infty} dq \rho^{\rm imp}_h (q)s(\la - g(q));\cr\cr
\sigma'^{\rm imp}_{pn}(\la) +\sigma'^{\rm imp}_{hn}(\la) &=&
\delta_{n1} \int^\infty_{-\infty} dq \rho^{\rm imp}_p (q)s(\la - g(q))\cr\cr
&& \hskip -1.25in + \int^\infty_{-\infty}d\la '
s(\la - \la')(\sigma'^{\rm imp}_{hn+1}(\la ')
+\sigma'^{\rm imp}_{hn-1}(\la ') )
\end{eqnarray}
In this form, the equations are easily solved numerically through iteration.

This computation of the finite temperature conductance comes with some potential pitfalls as discussed in
Ref. (\onlinecite{long}).  However as shown in this same reference we are able to accurately reproduce the universal
finite temperature linear response conductance scaling curve for a single dot.  Uncertainties arose
with the application of this methodology because of uncertainties with identifying the correct bulk scattering
states.  Here however the choice of bulk scattering states is exactly the same -- differences between the treatment
here and Ref. (\onlinecite{long}) only arise at the level of impurity scattering.  As our choice of bulk scattering
states was unproblematic previously, we expect it to be similarly so here.

\section{Analysis of Equations Governing Zero Temperature Transport of Double Dots in Parallel}

In this section we will analyze in detail the equations governing 
zero temperature transport in double quantum dots.
As discussed in Section II.B, there are two sets of constraints under which a pair of dots in parallel
is integrable.  The most interesting set of constraints is 
given in Eqn. (\ref{eIIix}):
\begin{eqnarray}\label{eIVi}
U_{\alpha\alpha '} &=& \delta_{\alpha \alpha'}U_{\alpha};~~U_{\alpha} \Gamma_\alpha = 
U_{\alpha'}\Gamma_{\alpha'};\cr\cr
&& \hskip -.14in U_{\alpha} + 2\epsilon_{d\alpha} = U_{\alpha'} + 2\epsilon_{d\alpha'};
\end{eqnarray}
These conditions correspond to well separated parallel dots.
They are neither capacitively nor tunneled coupled.  
The second set of conditions, corresponding to the constraints,
\begin{eqnarray}\label{eIVii}
U_{\alpha\alpha'} &=& U;~~\Gamma_\alpha = \Gamma_{\alpha'};~~\epsilon_\alpha = \epsilon_{\alpha'},
\end{eqnarray}
leads to transport that is identical in nature to a single level dot.  As such we will focus primarily in this section
on parallel dots meeting the first set of constraints and only briefly discuss dots 
constrained by the second set at the section's end.

At zero temperature, as discussed in Section III.A, the scattering phases which determines
the transport are expressed in terms of the two impurity
densities, $\rho^{\rm imp}(q)$ and $\sigma^{\rm imp}(\lambda)$.   These two densities satisfy
the two integral equations (Eqn. (3.7)):
\begin{eqnarray}\label{eIViii}
\rho^{\rm imp} (q) &=& \rho_0^{\rm imp} (q) + 
g'(q) \il a_1(g(q)-\la) \sig^{\rm imp} (\la); \cr\cr
\sig^{\rm imp} (\la ) &=& \sigma_0^{\rm imp}(\lambda)
- \ilp a_2(\la '-\la)\sig^{\rm imp} (\la ')\cr\cr
&& - \int^B_{-D}dq a_1(\la - g(q))\rho^{\rm imp} (q).
\end{eqnarray}
The `Fermi surfaces' of the dots, $Q$ and $B$ are determined by 
the conditions given in Eqn. (2.19).
As $Q$ and $B$ (up to $1/L$ corrections) are given solely by the bulk densities
$\sigma^{\rm bulk}(\lambda)$ and $\rho^{\rm bulk}(q)$, their behaviour, as a function
of the two quantities $2\epsilon_{d1/2}-U_{11/22}$ and $\Gamma_{1/2}U_{11/22}$, is exactly
the same as that of a single level dot and so discussed extensively in Refs.(\onlinecite{wie}) and (\onlinecite{long}).

The equations in Eqn. (\ref{eIViii}) can only be solved in general numerically.  However they do admit analytic solutions 
in certain cases in 
two limits: a) $|\epsilon_{d1}-\epsilon_{d2}| \gg \Gamma_1,\Gamma_2$ and b)
$|\epsilon_{d1}-\epsilon_{d2}| \ll \Gamma_1,\Gamma_2$.
In both cases the source terms, $\rho^{0\rm imp}(q)$ and $\sigma^{0\rm imp}(\lambda)$ can be written in the form
\begin{eqnarray}\label{eIViv}
\rho_0^{\rm imp}(q) &\approx& \Delta_{01}^{\rm imp}(q) + \Delta_{02}^{\rm imp}(q);\cr\cr
\sigma_0^{\rm imp}(\lambda) &\approx& \sigma_{01}^{\rm imp}(\lambda ) + \sigma_{02}^{\rm imp}(\lambda ),
\end{eqnarray}
where
\begin{eqnarray}\label{eIVv}
\Delta_{0i}^{\rm imp}(q) &=& \frac{\tilde\Gamma_i}{\tilde\Gamma_i^2+(q-\tilde\epsilon_{di})^2};\cr\cr
\sigma_{0i}^{\rm imp}(\lambda) &=& -{\rm Re}\frac{1}{\pi}\partial_\lambda\Delta_{0i}(x(\lambda)+iy(\lambda));\cr\cr
&=& \int^\infty_{-\infty} dq\Delta_{0i}(q)a_1(\lambda - g(q)) \cr
&& \hskip -.5in - a_1(\lambda-g(\tilde\epsilon_{di}+i\tilde\Gamma_i)) + 
\beta_ia_1(\beta_i\lambda-\gamma_i);\cr\cr
\alpha_i &=& \frac{U_i}{2} + \epsilon_{di}-\tilde\epsilon_{di};\cr\cr
\beta_i &=& \frac{U_i\Gamma_i}{2\alpha_i\tilde\Gamma_i}.
\end{eqnarray}
The parameters $\tilde\epsilon_{di}$ and $\tilde\Gamma_i$ are case dependent.  For
case a), we obtain:
\begin{eqnarray}\label{eIVvi}
&& \hskip -.98in {\bf |\epsilon_{d1}-\epsilon_{d2}| \gg \Gamma_1,\Gamma_2:} \cr\cr
\tilde\Gamma_{1/2} &=& \Gamma_{1/2} + \frac{\Gamma_{1}\Gamma_{2}}{(\epsilon_{d1}-\epsilon_{d2})^2}(\Gamma_{1/2}-\Gamma_{2/1});\cr\cr
\tilde\epsilon_{d1/2} &=& \epsilon_{d1/2} - \frac{\Gamma_{1}\Gamma_{2}}{\epsilon_{d1/2}-\epsilon_{d2/1}},
\end{eqnarray}
and for case b)
\begin{eqnarray}\label{eIVvii}
&& \hskip -1.17in {\bf |\epsilon_{d1}-\epsilon_{d2}|} \ll \Gamma_1,\Gamma_2: \cr\cr
\tilde\Gamma_{1} &=& \Gamma_{1}+\Gamma_{2};\cr\cr
\tilde\epsilon_{d1} &=& \frac{\epsilon_{d1}+\epsilon_{d2}}{2};\cr\cr
\tilde\Gamma_{2} &=& \frac{(\epsilon_{d1}-\epsilon_{d2})^2\Gamma_{1}\Gamma_{2}}{(\Gamma_1+\Gamma_2)^3};\cr\cr
\tilde\epsilon_{d2} &=& \epsilon_{d2} + (\epsilon_{d1}-\epsilon_{d2})\frac{\Gamma_2}{\Gamma_1+\Gamma_2}.
\end{eqnarray}
If $\epsilon_{d1}=\epsilon_{d2}$ exactly, then $\tilde\Gamma_2$ vanishes and $\Delta_{02}(q)$ is identically
zero.

Given the division of the source terms into two pieces and the linearity of the equations in (\ref{eIViii}), we can
correspondingly write $\sigma(\lambda) = \sigma_1(\lambda) + \sigma_2(\lambda)$ and 
$\rho(\lambda) = \rho_1(\lambda) + \rho_2(\lambda)$ where 
\begin{eqnarray}\label{eIVviii}
\rho_i^{\rm imp} (q) &=& \rho_{0i}^{\rm imp} (q) + 
g'(q) \il a_1(g(q)-\la) \sig_i^{\rm imp} (\la); \cr\cr
\sig_i^{\rm imp} (\la ) &=& \sigma_{0i}^{\rm imp}(\lambda)
- \ilp a_2(\la '-\la)\sig_i^{\rm imp} (\la ')\cr\cr
&& - \int^B_{-D}dq a_1(\la - g(q))\rho_i^{\rm imp} (q),
\end{eqnarray}

\subsection{Linear Response Conductance at Zero Magnetic Field}
We compute the Fermi surface scattering phase, $\delta_e$, governing the linear response conductance,
$G = 2e^2/h\sin^2(\delta_e/2)$, 
via
\begin{eqnarray}\label{eIVix}
\delta_e &=& 2\pi\int^{\infty}_{Q}(\sigma_1^{\rm imp}+\sigma_2^{\rm imp})(\lambda)\cr\cr
&=& 2\pi - \pi\int^Q_{-\infty}(\sigma_1^{\rm imp}+\sigma_2^{\rm imp})(\lambda)\cr\cr
&=& 2\pi - \delta_{e1} - \delta_{e2}.
\end{eqnarray}
The quantity, $\int^\infty_Q$, in first line of Eqn. (\ref{eIVix}) is the impurity particle number
while the corresponding integrated quantity in the second line is the impurity hole number.
$\sigma^{\rm imp}_i$ can be obtained via a Weiner-Hopf analysis of Eqn. (4.8) (with $B=-D$).  Our separation of the impurity
densities into two pieces allows the computation to proceed along the lines of Ref. (\onlinecite{wie}).
We relegate the analysis to Appendix A simply reporting the results for $\delta_{ei}$ below.  These
results apply to both $\delta_{e1}$ and $\delta_{e2}$ provided $\epsilon_{d1}\neq\epsilon_{d2}$.
If instead $\epsilon_{d1}=\epsilon_{d2}$, $\delta_{e2}$ is identically zero and the equations below
apply only to $\delta_{e1}$.

The analysis of $\delta_{ei}$ breaks down into three cases depending on the value of $Q$, the parameter describing the
position of the Fermi surface of the $\lambda$-excitations.
\vskip .1in
\noindent {\bf Case i: Dots Near the Particle-Hole Symmetric Point $Q < 0$}:  
Near the symmetric point, i.e. $U_i/2+\epsilon_{di} \ll \sqrt{U\Gamma}$,
 then $Q <0$.  Setting $J_i\equiv I_i^{-1}-Q$, where 
\begin{equation}\label{eIVixa}
I_i^{-1}=\frac{\alpha_i^2-\tilde\Gamma_i^2}{2U_i\Gamma_i},
\end{equation}
$\delta_{ei}$ is then given by
\begin{eqnarray}\label{eIVx}
\delta_{ei} &=& K_{i1} + K_{i2}.
\end{eqnarray}
Here $K_{ij}$ is the contribution to the scattering phase, $\delta_{ei}$, coming from the bare impurity densities,
$\Delta^{\rm imp}_{0j}(q)$ and $\sigma^{\rm imp}_{0j}(\lambda)$.
In this case we find (see Appendix A)
\begin{eqnarray}\label{eIVxi}
K_{i1} &=& \sqrt{\pi}\sum^{\infty}_{n=0} \frac{(-1)^n}{n+1/2}\bigg(\frac{n+1/2}{e}\bigg)^{n+1/2}\frac{1}{\Gamma (n+1)}\cr\cr
&& \hskip .25in \times \int^\infty_{-\infty} dq \Delta_{0i}(q) e^{-\pi (2n+1)(g(q)-Q)};\cr\cr
K_{i2} &=& {\rm sign}(1-\beta_i)\sqrt{\pi} {\bf P}\int^\infty_0 \frac{d\omega}{\omega}\bigg(\frac{\omega}{e}\bigg)^\omega
\frac{e^{-2\pi\omega J_i}}{\Gamma(1/2+\omega)}\cr\cr
&& \hskip .25in \times \frac{\sin (2\pi\omega(\gamma_i-1/2))}{\cos (\pi\omega)}\cr\cr
&&\hskip -.35in +~{\rm sign}(1-\beta_i)\sqrt{\pi}\sum^\infty_{n=0}
\frac{(-1)^n}{n+1/2}\bigg(\frac{n+1/2}{e}\bigg)\frac{e^{-(2n+1)\pi J_i}}{\Gamma(n+1)}\cr\cr
&& \hskip .35in \times \cos\big(\pi(\gamma_i-1/2)(2n+1)\big),
\end{eqnarray}
where ${\bf P}$ indicates that the principal value of the integral is to be taken and 
$\gamma_i =  1/2(1-\beta_i^{-1}){\rm sign}(1-\beta_i)$.  The value of the Fermi
surface, $Q$, was found in Refs. (\onlinecite{wie}) and (\onlinecite{long}) to be determined implicitly by
\begin{eqnarray}\label{eIVxii}
\frac{2\epsilon_{di}+U_i}{\sqrt{2U_i\Gamma_i}}\!\!\! &=&\!\!\!
\frac{\sqrt{2}}{\pi}\sum^\infty_{n=0}
\frac{(-1)^ne^{\pi Q(2n+1)}}{(2n+1)^{3/2}}G^+(i\pi(2n+1)).\cr
&&
\end{eqnarray}
\vskip .1in
\noindent {\bf Case ii: Mixed Valence Regime of the Dots: $Q>0$ and $J_i = I_i^{-1} - Q >  0$}:  
For these values of the parameters, we are in the mixed
valence regime of the dots.
Here $\delta_{ei}$ is given by
\begin{eqnarray}\label{eIVxiii}
\delta_{ei} &=& K_{i1} + K_{i2}\cr\cr
K_{i1} &=& \pi\bigg(\sqrt{2}-1-\frac{\pi}{\sqrt{2}6}J_i + \pi^2\frac{\sqrt{2}}{24^2}J_i^2  \cr\cr
&+& \frac{1}{\pi\sqrt{2}}\big(\beta_i^{-1}-1\big)J_i + {\cal O}(J_i^3) + {\cal O}(J_i(\beta_i^{-1}-1))\bigg);\cr\cr
K_{i2} &=& {\rm sign}(1-\beta_i)\sqrt{\pi} {\bf P}\int^\infty_0 \frac{d\omega}{\omega}\bigg(\frac{\omega}{e}\bigg)^\omega
\frac{e^{-2\pi\omega J_i}}{\Gamma(1/2+\omega)}\cr\cr
&& \hskip .25in \times \frac{\sin (2\pi\omega(\gamma_i-1/2))}{\cos (\pi\omega)}\cr\cr
&& \hskip -.35in +~{\rm sign}(1-\beta_i)\sqrt{\pi}\sum^\infty_{n=0}\frac{(-1)^n}{n+1/2}\bigg(\frac{n+1/2}{e}\bigg)\frac{e^{-(2n+1)\pi J_i}}{\Gamma(n+1)}\cr\cr
&& \hskip .35in \times \cos\big(\pi(\gamma_i-1/2)(2n+1)\big).
\end{eqnarray}
If $J_i$ is small, $K_{i2}$ can be approximated by
\begin{eqnarray}\label{eIVxiv}
K_{i2} &=& -{\rm sign}(1-\beta_i)\frac{\sqrt{2}}{\pi}\tan^{-1}\bigg(\frac{J_i}{\gamma_i}\bigg)
\end{eqnarray}
For $\gamma_i$ small (as would be the case if $|\epsilon_{d1}-\epsilon_{d2}| \ll \Gamma_{1,2}$)
we see that $K_{i2}$ undergoes a rapid variation as $J_i$ is varied.  This in turn will imply a rapid
variation in both the $T=0$ linear response conductance and the number of localized electrons in
the system as the gate voltage is swept.

In this case $Q$ needs to be determined by a self-consistent numerical evaluation of Eqns. (\ref{eIIxxii}).
If $Q \sim 1$ however, it is a reasonable approximation to take
\begin{equation}\label{eIVxv}
Q = q^* + \frac{1}{2\pi}\ln (2\pi e q^*),
\end{equation}
where $q^* = (\epsilon_{di}+U_i/2)^2/(2U_i\Gamma_i)$.
\vskip .1in

\noindent {\bf Case iii: Empty orbital regime: $Q > 0$ and $J_i = I_i^{-1} - Q < 0$}.  
In this parameter range, the dot filling is small.  $\delta_{ei}$ is then given by
\begin{eqnarray}\label{eIVxvi}
\delta_{ei} &=& K_{i1} + K_{i2}\cr\cr
K_{i1} &=& \pi + \frac{1}{\sqrt{\pi}}
\int^\infty_0 \frac{d\omega}{\omega}\Gamma(1/2+\omega)\bigg(\frac{e}{\omega}\bigg)^{\omega}e^{2\pi\omega J_i}\cr\cr
&& \hskip -.15in \times \sin(\pi\omega(1+\frac{1}{\beta_i}));\cr\cr
K_{i2} &=& \pi{\rm sign}(1-\beta_i) + \frac{1}{\sqrt{\pi}}
\int^\infty_0 \frac{d\omega}{\omega}\Gamma(1/2+\omega)\cr\cr
&& \hskip -.15in \times \bigg(\frac{e}{\omega}\bigg)^{\omega}e^{2\pi\omega J_i}
\sin(\pi\omega\gamma_i).
\end{eqnarray}
If $J_i$ is small, the expression for $K_{i2}$ in Eqn. (\ref{eIVxiv}) holds equally well.
Similarly, in this regime, $Q$ is given as above in Eqn. (\ref{eIVxv}).

\subsection{Finite Field Transport}

In finite field, we must compute the scattering phases, $\delta_{e\uparrow/\downarrow}$, separately
for spin up/down electrons.  Like at $H=0$, these scattering phases are given in terms of
the impurity densities:
\begin{eqnarray}\label{eIVxvii}
\delta_{e\uparrow} &=& 2\pi\int^{B}_{-D}dq(\rho^{\rm imp}_1(q) + \rho^{\rm imp}_2(q))  \cr
&& \hskip .5in + 2\pi\int^{\infty}_{Q}(\sigma_1^{\rm imp}+\sigma_2^{\rm imp})(\lambda);\cr\cr
\delta_{e\downarrow} &=& 2\pi\int^{\infty}_{Q}(\sigma_1^{\rm imp}+\sigma_2^{\rm imp})(\lambda).
\end{eqnarray}

In general these equations cannot be solved analytically.  However at the symmetric point they do admit such a solution.
At this point,\cite{long} $Q = \infty$ and $\sigma_i$ can be expressed solely in terms of
$\rho_i$ (simply by inverting Eqn. 4.8 via Fourier transform):
\begin{eqnarray}\label{eIVxviii}
\sigma^{\rm imp}_i(\lambda) &=& \int^\infty_{-\infty} d\lambda' (1+a_2)^{-1}(\lambda-\lambda') \sigma^{\rm imp}_{0i}(\lambda ')\cr
&& \hskip .2in -\int^B_{-D} dq \rho^{\rm imp}_i (q) s(\lambda-g(q)).
\end{eqnarray}
In turn, we obtain an equation for $\rho^{\rm imp}_i$ independent of $\sigma^{\rm imp}_i$:
\begin{eqnarray}\label{eIVxix}
\rho^{\rm imp}_i (q) \!\!&=&\!\! \rho^{\rm imp}_{0i}(q)\!-\! g'(q)\int^B_{-\infty}\!\!dq' 
R(g(q) - g(q'))\rho^{\rm imp}(q'),\cr
&&
\end{eqnarray}
where 
\begin{equation}\label{eIVxx}
R(\lambda) = \frac{1}{2\pi}\int^\infty_{-\infty}d\omega \frac{e^{-i\omega\lambda}}{1+e^{|\omega|}}.
\end{equation}
and we can write $\rho^{\rm imp}_{0i}(q)$ as
\begin{eqnarray}\label{eIVxxi}
\rho^{\rm imp}_{0i}(q) &=& \Delta_{0i}(q) + g'(q)\int^\infty_{-\infty}dq R(g(q)-g(q'))\Delta_{0i}(q)\cr\cr
&& \hskip -.6in + {\rm sign}(1-\beta_i)g'(q)\int^\infty_{-\infty}\frac{d\omega}{2\pi}
e^{-i\omega(g(q)-I_i^{-1})}\cr\cr
&&\hskip .8in \times \frac{e^{\frac{|\omega|}{2}(1-\beta_i^{-1}){\rm sign}(1-\beta_i)}}{2\cosh(\frac{\omega}{2})}.
\end{eqnarray}
We are most interested in the behaviour of $\rho^{\rm imp}_i$ for $q<0$ as this is the region relevant for
magnetotransport when $H$ is on the order of the Kondo temperature.
Following Ref. (\onlinecite{wie}), for $q<0$ the first two terms may be rewritten as
\begin{eqnarray}\label{eIVxxii}
\Delta_{0i}(q) &+& g'(q)\int^\infty_{-\infty}dq R(g(q)-g(q'))\Delta_{0i}(q) = \cr\cr
&&\hskip -.4in -
g'(q)\int^\infty_{-\infty}\frac{d\omega}{2\pi}\frac{e^{-i\omega(g(q)-I_i^{-1})+\frac{|\omega|}{2}{1-\beta^{-1}}}}
{2\cosh(\frac{\omega}{2})}.
\end{eqnarray}
At the symmetric point, the scattering phases similarly simplify reducing to
\begin{eqnarray}\label{eIVxxiii}
\delta_{e\uparrow} &=& 2\pi + \pi\int^{B}_{-D}dq(\rho^{\rm imp}_1(q) + \rho^{\rm imp}_2(q));\cr\cr
\delta_{e\downarrow} &=& 2\pi - \pi\int^{B}_{-D}dq(\rho^{\rm imp}_1(q) + \rho^{\rm imp}_2(q)).
\end{eqnarray}
We thus see that even though $\delta_{e\uparrow}$ differs from $\delta_{e\downarrow}$, the conductance
for each spin species is the same.

With the source term in this form, Eqn. (\ref{eIVxxii}) can be solved by a Wiener-Hopf analysis.  
We again relegate the details to Appendix A and
here simply report the results.
Firstly we consider the dependence of the limit $B$, the Fermi surface of the q-excitations, on the
magnetic field.  It has the same functional dependence 
on $U_i+2\epsilon_{di}$ and $\Gamma_iU_i$ as in the case of a single level dot.  Thus we
can simply borrow results from Refs. (\onlinecite{wie}) and (\onlinecite{long}):
\begin{eqnarray}\label{eIVxxiv}
\frac{H}{2\pi} &=& \bigg(\frac{U_i\Gamma_i}{8\pi^2}\bigg)^{1/2}\int^\infty_{-\infty}d\omega 
e^{-i\omega \frac{B^2}{2U_i\Gamma_i}}\frac{1}{(i\omega+\delta)^{1/2}}\frac{1}{\omega-i\delta}\cr\cr
&& \hskip .7in 
\times \frac{1}{\Gamma(\frac{1}{2}+\frac{i\omega}{2\pi})}\bigg(\frac{i\omega+\delta}{2\pi e}\bigg)^{\frac{i\omega}{2\pi}}.
\end{eqnarray}
In the small $H$ limit, one can deform the contour into the lower half plane,
merely taking into account the pole nearest zero.  We so obtain 
\begin{equation}\label{eIVxxv}
b \equiv \frac{B^2}{2U_i\Gamma_i} = -\frac{1}{2\pi}\log\bigg(\frac{\pi e H^2}{4 U\Gamma}\bigg).
\end{equation}
Next we write down expressions for $\int^{B}_{-D}dq\rho^{\rm imp}_i(q)$:
\begin{equation}\label{eIVxxvi}
\int^{B}_{-D}dq\rho^{\rm imp}_i(q) = S^{\rm Kondo}_i + S^{\rm charge}_i.
\end{equation}
Here $S^{\rm Kondo}_i$ encodes information that varies on the Kondo scale.  It is given by
\begin{widetext}
\begin{eqnarray}\label{eIVxxvii}
S^{\rm Kondo}_i = 
\begin{cases}
\frac{2}{\sqrt{\pi}} \Theta(\beta_i-1)
\sum^\infty_{n=0}\frac{(-1)^n}{\Gamma(1+n)(n+\frac{1}{2})}\bigg(\frac{H\sqrt{\pi e}}{T^{RKKY}_K 2^{3/2}}\bigg)^{2n+1}
\bigg(\frac{n+\frac{1}{2}}{e}\bigg)^{n+\frac{1}{2}}
\cos(\pi(n+\frac{1}{2})(1-\beta_i^{-1})),
& \text{if $b > I_i^{-1}$,} \cr
2\Theta(\beta_i-1)\bigg[ 1 - \frac{1}{\pi^{3/2}}\int^\infty_0 \frac{d\omega}{\omega} \bigg(\frac{8 (T^{RKKY}_K)^2}{\pi\omega H^2}\bigg)^\omega
\Gamma(\frac{1}{2}+\omega)\sin(\pi\omega)\cos(\pi\omega(1-\beta_i^{-1}))\bigg], & \text{if $b < I_i^{-1}$.}\cr
\end{cases}
\end{eqnarray}
\end{widetext}
where the Kondo temperature is defined by
\begin{equation}\label{eIVxxviii}
T^{RKKY}_K = \bigg(\frac{U_i\Gamma_i}{2}\bigg)^{1/2}e^{-\pi I_i^{-1}}, ~~~\text{if $\beta_i > 1$}.
\end{equation}
(We explain the label 'RKKY' in Section V.)
This definition is well posed as of $\beta_1$ and $\beta_2$, only one will be ever greater than 1.
This Kondo temperature is similar to what is found in a single level dot but for the form
of $I_i^{-1}$ (see the definition in Eqn. (\ref{eIVixa}) involving 
parameters renormalized by the effects of interference
between the two dots).  We can evaluate $S^{\rm Kondo}_i$ in the limit $H \gg T^{\rm RKKY}_K$ 
(but $H < \sqrt{2U_i\Gamma_i/\pi e}$):
\begin{eqnarray}\label{eIVxxix}
S^{\rm Kondo}_i &=& 2\Theta(\beta_i-1)\bigg[1 - \frac{1}{2\log(H/\tilde T^{RKKY}_K)} \cr\cr
&& \hskip -.5in + 
{\cal O}\bigg(\frac{\log\log(\htk)}{\log^2(H/\tilde T^{RKKY}_K)}\bigg) + 
\frac{\pi^2(1-\beta_i^{-1})^2}{8\log^3(H/\tilde T^{RKKY}_K)} \cr\cr
&& \hskip -.5in +
{\cal O}\bigg(\frac{(1-\beta_i^{-1})^2\log\log(\htk)}{\log^4(H/\tilde T^{RKKY}_K)}\bigg) \cr\cr
&& \hskip -.5in +
{\cal O}\bigg(\frac{(1-\beta_i^{-1})^4}{\log^5(H/\tilde T^{RKKY}_K)}\bigg)\bigg],
\end{eqnarray}
where $\tilde T^{\rm RKKY}_K = \frac{2^{3/2}}{\sqrt{\pi e}}T^{\rm RKKY}_K$.
We have included the leading order term in $(1-\beta_i^{-1})^2$ as this governs 
the differing behaviour of $S^{\rm Kondo}_i$ in the two cases of $\delep \ll \Gamma_{1,2}$
and $\delep \gg \Gamma_{1,2}$.

The second term, $S^{\rm charge}$, varies on a scale comparable to the charging energy of the dot, $\sqrt{U_i\Gamma_i}$.
In the dots' Kondo regime this scale will be much larger than $T_K$.  It is given by
\begin{eqnarray}\label{eIVxxx}
S^{\rm charge}_i &=&
\frac{1}{\sqrt{\pi}}\sum^\infty_{n=0}\frac{e^{-2\pi b(n+\frac{1}{2})}}{\Gamma(1+n)(n+\frac{1}{2})}
\bigg(\frac{n+\frac{1}{2}}{e}\bigg)^{n+\frac{1}{2}}\cr\cr
&& \hskip -.5in
\times \int^\infty_{-\infty}dq e^{-2\pi g(q)(n+\frac{1}{2})}\Delta_{0i}(iq),
\end{eqnarray}
and for fields $H \sim \tk$ makes a far smaller contribution than that of $S^{\rm Kondo}$.

\subsection{Two degenerate dots in parallel}

We now briefly consider two degenerate dots in parallel satisfying the conditions given
in Eqn. (\ref{eIVii}).  This case is less interesting than the one described by
the conditions in Eqn. (\ref{eIVi}) as it is equivalent to a single level dot with
Coulomb repulsion, $U$, and linewidth $\Gamma = \sum_\alpha \Gamma_\alpha$.  This follows
directly from Eqns. (\ref{eIIv}) and (\ref{eIIxvi}), the equations
governing the bare impurity scattering phase, $\delta(q)$, and the function, $g(q)$, controlling the integrability
of the dot model.  Because $U_{\alpha\alpha'} = U$ and
$\epsilon_{d\alpha}=\epsilon_d$, we can make the replacement $\Gamma \rightarrow \sum_\alpha \Gamma_\alpha$
and obtain equations corresponding to a single level dot.

With this equivalence, we can directly apply the analysis of the linear response conductance contained
in Ref. (\onlinecite{long}).  And so none of interesting phenomena identified in the coming two sections
remains.  In particular the RKKY Kondo effect, the presence of interference phenomena, and a non-trivial
form of the Friedel sum rule are all absent.

\section{Examples of Zero Temperature Transport of Double Dots in Parallel}

In this section we consider various aspects of zero temperature transport in double
quantum dots both in the absence and presence of a magnetic field.  We will see that
there is a mixture of different types of Kondo physics with interference effects.  We will also
will see that the number of electrons displaced in attaching the quantum dots to the leads can
be negative, a contrast to what is found in single level quantum dots.
We begin by examining $H=0$ conductance of the dot system as a function of gate voltage.

\subsection{Zero Field Linear Response Conductance as a Function of Gate Voltage}

\begin{figure*}[tbh]
\centerline{\psfig{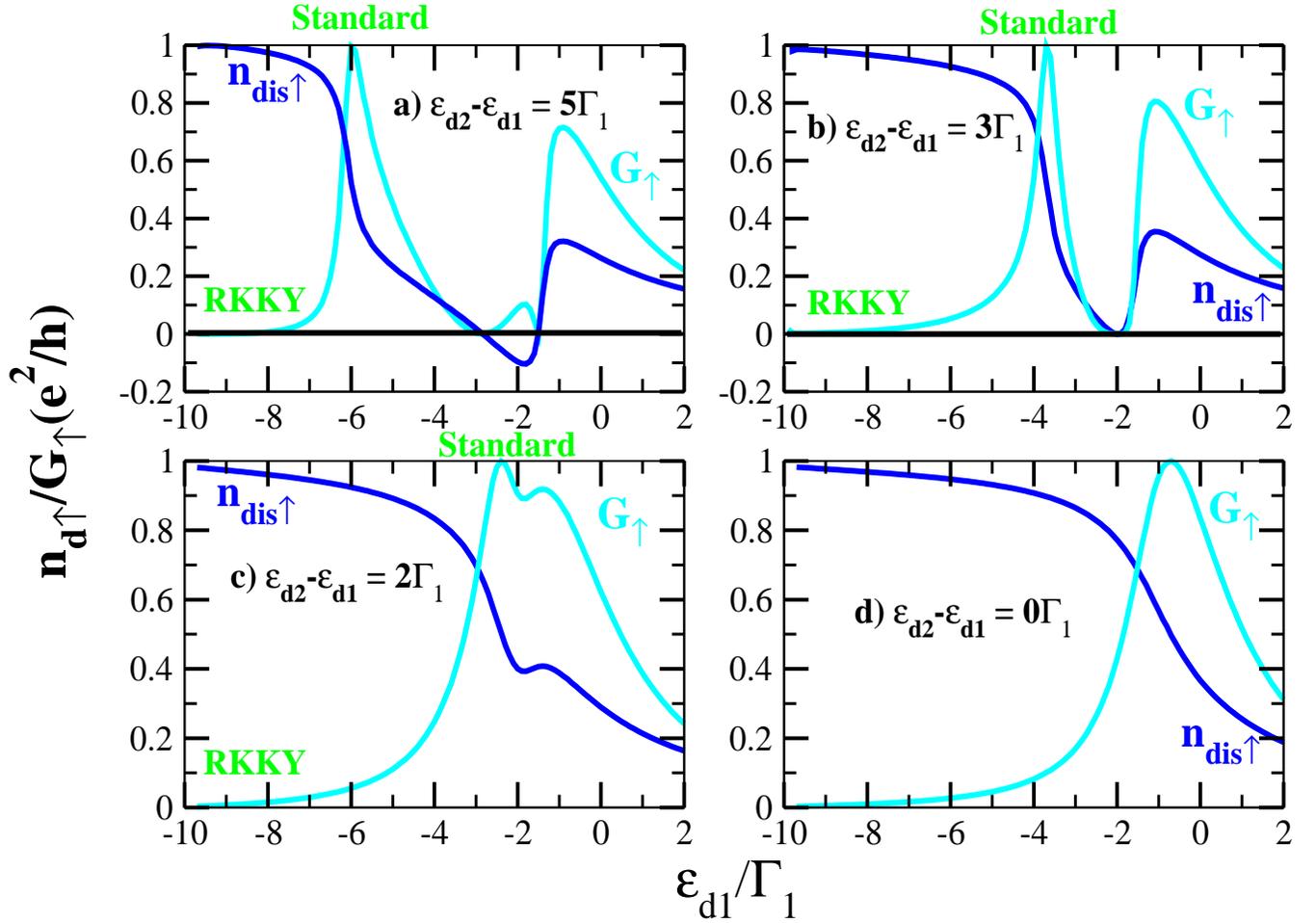}}
\caption{Plots of the T=0 H=0 linear response conductance as a function of
gate voltage.  In making these plots, we chose $U_1=1$ and $\Gamma_1 = U_1/20$
uniformly and values of $|\epsilon_{d1}-\epsilon_{d2}|$ to be $5\Gamma_1$,
$3\Gamma_1$, $2\Gamma_1$ and $0$ in panels a, b, c, and d.
$\Gamma_2$ and $U_2$ are then chosen in accordance with the integrability conditions of
Eqn. (\ref{eIIix}).  On the plots we mark the values of gate voltage where the RKKY Kondo
effect and the standard Kondo effect take place.}
\end{figure*}

Here we study the behaviour of the $T=0$ linear response conductance, $G$, as a function of gate voltage
keeping the distance between the dot levels, i.e. $|\epsilon_{d1}-\epsilon_{d2}|$, fixed.  Thus we
allow the gate voltage to move the two dot chemical potentials in unison.  In Figure 3 we plot
four examples of $G$ for dot systems with differing separations ranging from
$|\epsilon_{d1}-\epsilon_{d2}| >> \Gamma_1, \Gamma_2$ to $|\epsilon_{d1}-\epsilon_{d2}|=0$.
Provided that $|\epsilon_{d1}-\epsilon_{d2}| \neq 0$, the conductance exhibits three
basic features: i) an RKKY Kondo effect at the particle-hole symmetric point of the dot;
ii) a standard Kondo effect corresponding to a value of gate voltage when one electron sits on
the dot; and iii) interference effects whose strength is proportional to the separation
of the two bare dot levels.

To understand this better let us begin by consider any of first three panels of Figure 3 (a, b, or c).  Here 
$|\epsilon_{d1}-\epsilon_{d2}|$ is nonzero.  The linear response conductance
trace begins at the particle hole symmetric point of the dot where (uniformly) $\epsilon_{d1} = -10\Gamma_1$
and $\epsilon_{d2} = -5\Gamma_1, -3\Gamma_1, -2\Gamma_1$ respectively for panels a), b), and c).  
At this point exactly two electrons sit on the two
dot system.  By the Friedel sum rule the corresponding scattering phase is $\delta_e = 2\pi$
and the conductance vanishes.  Although the conductance vanishes as this point, and counter
to the intuition we have in a single level dot where Kondo physics is associated with
a unitary maximum in the conductance, Kondo physics
is still present.  We determine its presence by studying the low energy density of states 
of the impurity double dot system.  We find what amounts to an Abrikosov-Suhl resonance.

\begin{figure*}[tbh]
\centerline{\psfig{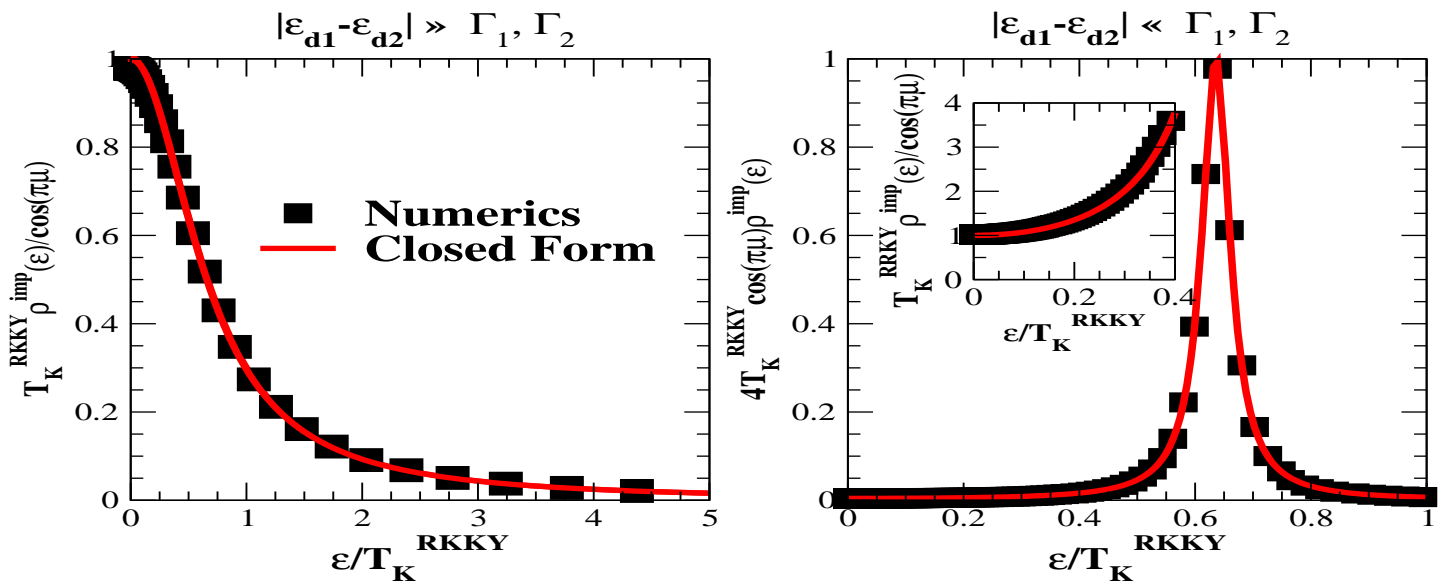}}
\caption{The low lying impurity density of states (the Abrikosov-Suhl resonance)
as a function of energy present in the double dot system at its particle-hole
symmetric point.  In the left panel we show the shape (Lorentzian) of the resonance
for double dots with well separated levels.  In the right panel we show the same
resonance for the case $\delep \ll \Gamma_{1,2}$ (but $\epo \neq \ept$).  Here
the resonance has split about zero energy by an amount $4\tk/\pi$.}
\end{figure*}

In the Bethe ansatz language this amounts to studying
the quantity $\rho^{\rm imp}(q(\epsilon))$ where $\rho^{\rm imp}(q)$ is the impurity
density of states of the charge excitations and $\epsilon_q(q)$ is their corresponding bulk
excitation energy\cite{long,kawa_den}.  In zero magnetic field, $\rho^{\rm imp}(q)$
is equivalent to $\rho^{\rm imp}_0(q)$
computed in Appendix (A3):
\begin{equation}\label{eVi}
\rho^{\rm imp}(q<0) = -2g'(q)\frac{\cos(\mu\pi)\cosh(\pi(g(q)-I^{-1}_i))}{\cosh(2\pi(g(q)-I^{-1}_i)) + \cos(2\pi\mu)}.
\end{equation}
where $g(q) = (q-\epsilon_{di}-U_i/2)^2/(2U_i\Gamma_i)$ and $\mu$ is given by
\begin{equation}\label{eVii}
\mu =
\begin{cases}
\frac{1}{2}(1-\beta_1^{-1}) & \text{if $\beta_1 > 1$};\cr
\frac{1}{2}(1-\beta_2^{-1}) & \text{if $\beta_2 > 1$},\cr
\end{cases}
\end{equation}
where $\beta_{1,2}$ are defined in Eqn. (\ref{eIVv}).
Given the equivalence of the bulk equations between the single level dot and 
our two dots in parallel, the bulk energy of charge excitations, $\epsilon_q(q)$, can be taken
from Refs. (\onlinecite{wie}) and (\onlinecite{long}):
\begin{equation}\label{eViii}
\epsilon_q (q) = \frac{\sqrt{2U_i\Gamma_i}}{\pi}e^{-\pi g(q)}.
\end{equation}
This relationship is valid provided $\epsilon_q(q)$ is on the order of the Kondo temperature.
Substituting Eqn.(\ref{eViii}) into Eqn. (\ref{eVi}) we obtain the form of the Abrikosov-Suhl
resonance of the dot system as a function of energy, $\ep$:
\begin{eqnarray}\label{eViv}
\rho^{\rm imp}(\epsilon) &=& \frac{\cos(\pi\mu)}{T^{RKKY}_K}\frac{1+\tilde\epsilon^2}
{2\tilde\epsilon^2\cos(2\pi\mu)+\tilde\epsilon^4+1},
\end{eqnarray}
where
$\tilde\epsilon = \frac{\pi\epsilon}{2T^{RKKY}_K}$. 
We plot the resonances for two cases, $\delep \gg \Gamma_{1,2}$ and $\delep \ll \Gamma_{1,2}$ in Figure 4.
The plot is for positive energies (as the Bethe ansatz solution only directly gives positive energy
information).  However as we are at the particle-hole symmetric point the resonance itself will be
symmetric about zero energy.

In the case $|\epsilon_{d1}-\epsilon_{d2}| \gg \Gamma_{1,2}$, i.e. $\mu \sim 0$, we obtain a resonance that
has the familiar Lorentzian form associated with a single dot.\cite{kawa_den,long} (see Figure 4).
However this resonance has twice the integrated weight in comparison
with a single dot.  Thus even though $|\epsilon_{d1}-\epsilon_{d2}| \gg \Gamma_{1,2}$, we cannot think
of this resonance as arising solely from Kondo physics involving the energetically lower of the two dot levels.

If $0 < |\epsilon_{d1}-\epsilon_{d2}| \ll \Gamma_{1,2}$, i.e. $\mu \sim 1/2$, 
the resonance dramatically changes structure.  Instead
of peaking at zero energy it has a sharp peak instead at $\epsilon = 2T_K^{RKKY}/\pi$.  While the resonance
does not vanish at zero energy in this case (inset to right panel of Figure 4), its value there is dramatically
smaller than at $\epsilon = 2T_K^{RKKY}/\pi$ (by a factor of $1/4\cos^2(\pi\mu$)).  
The resonance has thus split about zero energy.  This splitting is akin to the splitting seen
in the Kondo resonance when a interdot tunneling term, $d^\dagger_1d_2 + d^\dagger_2d_1$, 
is turned on between two degenerate dots, i.e. $\epo=\ept$.\cite{c1,c4,georges}  However
unlike this case the splitting is not proportional to the bare scale $\delep$, but rather to
the much smaller dynamical scale, $\tk$.

A note of caution is in order.  What we have just computed is the low energy density of states due the impurity
double dot.  But it is
not exactly equivalent to the impurity Green's function
$$
\langle d^\dagger_1d_1\rangle_{\rm ret}(\omega) + \langle d^\dagger_2 d_2\rangle_{\rm ret}(\omega)
$$
Formally the two differ by energy dependent matrix elements (matrix elements which are not directly computable
in the Bethe ansatz).  In the single dot case these matrix elements have, at best, only a weak dependence on
energy.  We would guess that this holds in the double dot case.  But we note that regardless of 
whether the two can be exactly
identified (up to a scaling constant), this assumption of equivalence 
does not affect the exactness of our $T=0$ results for transport.

\begin{figure}[tbh]
\centerline{\psfig{figure=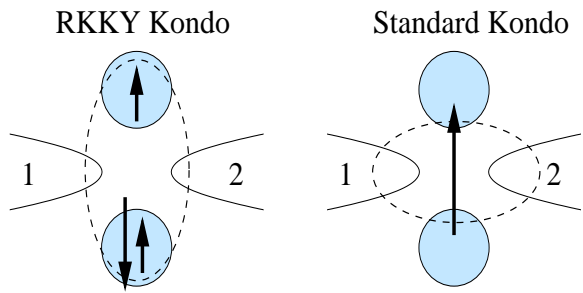,height=1.5in,width=3in}}
\caption{A cartoon of the two different types of Kondo effects that might appear in the
double dot: i) an RKKY Kondo effect where two electrons total sit on the dots and form a singlet
mediated through hopping on and off the leads -- as this effect depends
upon $\epsilon_{d1}\neq\epsilon_{d2}$, the dots possess an unequal distribution of
electron; and ii) the standard Kondo effect where a single
electron sits on the two dots localized primarily in the lower dot level.}
\end{figure}

We now turn to what sort of Kondo effect is present at the particle-hole symmetric point.
The Kondo singlet that forms at this point involves both electrons sitting on the two dots.
Because we know that an Abrikosov-Suhl resonance forms, we know the electrons are not
forming a direct singlet with one another (and indeed, in our model there is no direct interaction
between the dots).  Rather the singlet formed involving the two localized electrons is mediated by the conduction
electrons in the leads.  We thus term this Kondo effect an RKKY Kondo effect.  We sketch this type of Kondo effect in 
Figure 5.

It is clear that the RKKY physics occurring here is considerably different than the ferromagnetic
interaction that fourth order perturbation theory would predict between closely spaced dots.\cite{utsumi}
We see both that RKKY interaction is effectively antiferromagnetic and occurs at an energy scale akin
the Kondo temperature.  Moreover a strong ferromagnetic interaction would be characterized by a far different
conductance.  If the two electrons did form an effective spin-1 impurity, the $T=0$ conductance
would instead be $2e^2/h$ not the $0e^2/h$ we universally find.

The ground state associated with the RKKY Kondo effect is Fermi liquid (as will be obvious when we consider
the low temperature or low field conductances).  This contrasts with the non-Fermi liquid nature
of electrons bound into an effective spin-1.\cite{pos}  It is a check on our results that we do find Fermi
liquid physics.
We know that the properties of a single dot is perturbative {\it in the Coulomb repulsion} and so the dot's
low temperature fixed point is Fermi liquid.  The perturbative structure in the Coulomb repulsion
of a double dot is similar in nature.  There is then a predilection to believe that the low temperature physics of
the double dot should similarly be Fermi liquid.  
We find further support for our Fermi liquid picture in Ref. \onlinecite{jones}.  
While Ref. \onlinecite{jones} studied a two impurity Anderson model coupled to two electron channels (not one as here),
they found that the T=0 electron scattering always obeyed the Friedel sum rule, a signature of Fermi
liquid physics, no matter the anisotropy between the channels.

Moving away from the particle-hole symmetric point by adjusting $\epsilon_{d1,2}$ upwards (keeping their relative
distance fixed) we uniformly decrease the number of electrons on the dot until we have in total one
localized electron.  At this point we then expect a standard type Kondo effect to take place.  This
Kondo effect is akin to that seen in single level dots at their particle-hole symmetric point.  While
analyzing the Bethe ansatz equations in closed form is difficult, we can nonetheless guess what
the associated Kondo temperature is:
\begin{equation}\label{eVv}
T^{\rm Standard}_K = \bigg(\frac{U_i\Gamma_i}{2}\bigg)^{1/2}e^{\frac{\epsilon_{d1}(U_1+\epsilon_{d1})}{2U_1\Gamma_1}}.
\end{equation}
This formula, available from the analysis of single level dots\cite{wie,long,haldane,costi}
is valid provided $|\epsilon_{d1}-\epsilon_{d2}| \ll \Gamma_1,\Gamma_2$ and $\Gamma_1 \ll U_1$.
In our analysis of the magnetoconductance in the next section we verify that indeed $T^{\rm Standard}_K$ as given
above governs the physics.

When however $|\epsilon_{d1}-\epsilon_{d2}| \sim \Gamma_1,\Gamma_2$, the behaviour of the double dot system with
one localized electron appears Kondo like (as determined by the mangetoconductance) 
over a much narrower range of magnetic field (in units of the Kondo temperature).

As we further increase the gate voltage, we move into a regime characterized by a mixture of
interference influenced by strong correlations on the dots.  We see non-monotonic behaviour both in the
conductance and the number of electrons, $n_{\rm dis}$, {\it displaced} by the dots.  Moreover if 
$|\epsilon_{d1}-\epsilon_{d2}| \gg \Gamma_1,\Gamma_2$
we see that $n_{\rm dis}$ can go negative.  This requires further explanation.  

The Bethe ansatz does not allow one to direct compute the number of electrons, $n_d$, occupying the dots.
Rather it allows one to compute this number, $n_d$, plus $1/L$ changes of electron occupancy that
arise in coupling the leads to the dots.  Specifically
\begin{equation}\label{eVvi}
n_{\rm \sigma dis} = n_{d\sigma1} + n_{d\sigma2} + \int dx \bigg[\langle c^\dagger_{e\sigma}(x)c_{e\sigma}(x)\rangle - \rho_{\sigma\rm bulk }\bigg]
\end{equation}
where $\rho_{\sigma\rm bulk}$ is the density of electrons of spin $\sigma$
in the leads in the absence of any coupling to the dots.  
In terms of transport properties, this is a more natural quantity to compute.  
It is the quantity that is related to the scattering
phase by the Friedel sum rule, as the proof of Langreth explicitly shows.\cite{langreth}  We comment on this further
in the discussion section.

Now by definition $n_{d\sigma}$, the number of electrons on the dots, can never be negative.  Thus when we
find $n_{\sigma \rm dis}$ to be negative, we know that coupling the dots to the leads has induced a {\it negative}
$1/L$ correction to the lead electron density.

In this non-monotonic regime we see that sharp variations exist in the conductance\cite{c7}
and are most pronounced if $|\epsilon_{d1}-\epsilon_{d2}| \gg \Gamma_{1,2}$.
In particular, the variations occur
at two values of the gate
voltage (i.e. $\epsilon_{d1}$).
The first variation is associated with a sudden decrease in the number of localized electrons $n_{\rm \sigma dis}$
as $\epsilon_{d1}$ is lowered while the second variation sees a sudden increase in $n_{\rm \sigma dis}$.
We can characterize the two values of $\epsilon_{d1}$ where these variations occur.

The variations correspond to values of $\epsilon_{d1}$ where the contribution to the scattering phase
from $K_{i2}$ is rapidly varying (see Eqn. (\ref{eIVxiv}) of Section IV.A).  This rapid variation occurs
when $Q$, the parameter marking the Fermi surface of the $\lambda$-excitations, is close to $I_i^{-1}$.
Provided $Q>0$ we can employ Eqn. (\ref{eIVxv}) to parameterize $Q$ in terms of $\epsilon_{d1}$.
Equating $Q$ and $I_i^{-1}$ then leads to the two values of $\epsilon_{d1}$ (corresponding 
to the two values of $I_{1,2}^{-1}$) where the variation occurs
\begin{eqnarray}\label{eVvii}
\epsilon_{d1} \!\!\!&=&\!\!\! \bigg[\frac{2U_1\Gamma_1}{1+\frac{1}{2\pi I_i^{-1}}}\big(I_i^{-1} + \frac{1}{2\pi} - 
\frac{1}{2\pi}\log (2\pi e I_i^{-1})\big)\bigg]^{1/2} \!\!-\!\! \frac{U_1}{2}.\cr
&&
\end{eqnarray}

The sharpness of these variations in the conductance and $n_{\rm \sigma dis}$ are governed by the parameter 
$$
\gamma_i = \frac{1}{2}|1-\beta_i^{-1}|,
$$
a consequence of $K_{i2}$ behaving as $\tan ((I_i^{-1}-Q)/\gamma_i)$ for $I_i^{-1}-Q$ small
(see Eqn. (\ref{eIVxiv})).
$\gamma_i$ vanishes in the limit $|\epsilon_{d1}-\epsilon_{d2}| \rightarrow \infty$
while becoming large in the opposite limit.
As the two variations in $n_{\rm \sigma dis}$ are of the opposite sign, 
the first of the two variations will induce $n_{\rm \sigma dis}$ to go negative if it does not overlap
significantly with the second.  Thus the condition for $n_{\rm \sigma dis}$ to go negative
for some value of the gate voltage is
$$
|I_1^{-1}-I_2^{-1}| > \gamma_1 + \gamma_2
$$
that is, the spacing of the variations is larger than the sum of their widths.
We can translate this into a condition upon the necessary separation of the two levels.  We need
\begin{eqnarray}\label{eVviii}
|\epsilon_{d1}-\epsilon_{d2}| &>& \frac{b+\sqrt{b^2+4ca}}{2a}; \cr\cr
a &=& \frac{U_1^2-U_2^2}{4}; \cr\cr
b &=& \Gamma_1\Gamma_2(U_1+U_2);\cr\cr
c &=& 2U_i\Gamma_i|\Gamma_2^2-\Gamma_1^2|,
\end{eqnarray}
if $n_{\rm \sigma dis}$ is to go negative.
If $\Gamma_1,\Gamma_2 \ll U_1,U_2$ this condition simplifies to
\begin{eqnarray}\label{eVix}
|\epsilon_{d1}-\epsilon_{d2}| > \bigg(\frac{8U_i\Gamma_i|\Gamma_1^2-\Gamma_2^2|}{|U_1^2-U_2^2}\bigg)^{1/2}.
\end{eqnarray}
A note of caution is needed here.  This condition is derived under the constraints governing
integrability (Eqn. (\ref{eIVi})).  While self-consistent, it should not be used to determine
the minimum level separation in two dots with {\it arbitrary} parameters so as to guarantee
$n_{\rm \sigma dis}$ will be negative.  However one might nonetheless suppose that this condition
would provide a rough estimate of the minimum separation for such dots.  

Up to this point we have discussed in detail the behaviour of the conductance provided
$\epsilon_{d1}$ is not exactly equal to $\epsilon_{d2}$.  If instead
$\epsilon_{d1}=\epsilon_{d2}$, we do not see a marked change in the $T=0$, $H=0$ linear
response conductance.  With $\epsilon_{d1}$ close to $\epsilon_{d2}$, we already see in Figure
3d) that the rapid variations in $G$ associated with interference effects disappear.
(This is consistent with the slave boson study of Ref. (\onlinecite{c5}).)
And as $\epsilon_{d1}$ becomes equal to $\epsilon_{d2}$, the
functional dependence of $G$ upon the gate voltage remains the same (compare panels c) and
d) in Figure 3).  However at $\epo=\ept$ there is one significant change.  At the particle-hole
symmetric point, the RKKY Kondo effect disappears.
As this vanishing takes place {\it discontinuously}, $\epsilon_{d1}=\epsilon_{d2}$
marks a first order quantum critical point.
\vskip .1in
\begin{figure}[tbh]
\centerline{\psfig{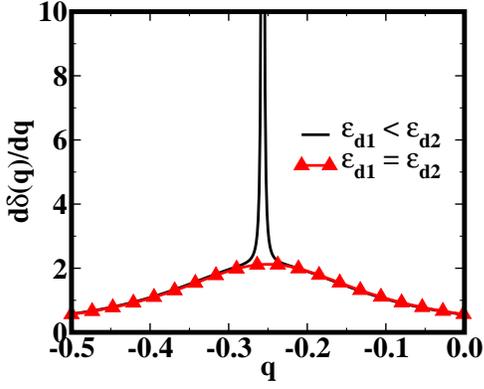}}
\caption{The impurity density of states for a non-interacting double dot
system when $\epsilon_{d1}$ and $\epsilon_{d2}$ are close but not
exactly equal and for when $\epsilon_{d1}=\epsilon_{d2}$. We see that
at $\epsilon_{d1} \rightarrow \epsilon_{d2}$, there is a discontinuous
change in the density of states.}
\end{figure}
\vskip .1in
This critical behaviour at $\epsilon_{d1}=\epsilon_{d2}$
has been observed before,\cite{goldstein,meden} and can be most easily understood
by first appealing to the non-interacting limit.  In the non-interacting limit, one can
diagonalize the dot degrees of freedom via the transformation,
$$
d_{even/odd} = \frac{1}{\sqrt{\Gamma_1^2+\Gamma_2^2}}(\Gamma_{1/2}d_1 \pm \Gamma_{2/1}d_2),
$$
and so we have a system where only one dot level is tied to the leads.\cite{gefen}
This mapping is tied to a dramatic change in the non-interacting density of states
(see Figure 6).  We see that with $\epsilon_1$ and $\epsilon_2$ nearly
equal but still different, the impurity density of states possesses a sharply peaked
feature not present when $\epsilon_1=\epsilon_2$.  It is this feature that
permits the formation of the Abrikosov-Suhl resonance when interactions are turned on.  
Its absence at  $\epsilon_1=\epsilon_2$ explains the absence of the RKKY Kondo effect.
While interactions
will couple the second (odd) dot level to the system, 
the absence of direct hopping to
the odd level means that the RKKY Kondo effect cannot develop.  Instead, the two
electrons bind directly into a singlet and the Kondo effect is entirely absent.

We point out that the particular behaviour observed at $\epsilon_{d1}=\epsilon_{d2}$ is dependent
on there being no coupling between the two dots (whether it be tunnel or capacitive coupling).
If the dots are coupled, the two levels will be split by the coupling (for example, a tunneling
term will lead to split bonding/antibonding dot levels).\cite{c1,c4}

\subsection{Finite Field Linear Response Conductance at the Dots' Particle-Hole Symmetric Point}

\begin{figure*}[tbh]
\vskip 1in
\centerline{\psfig{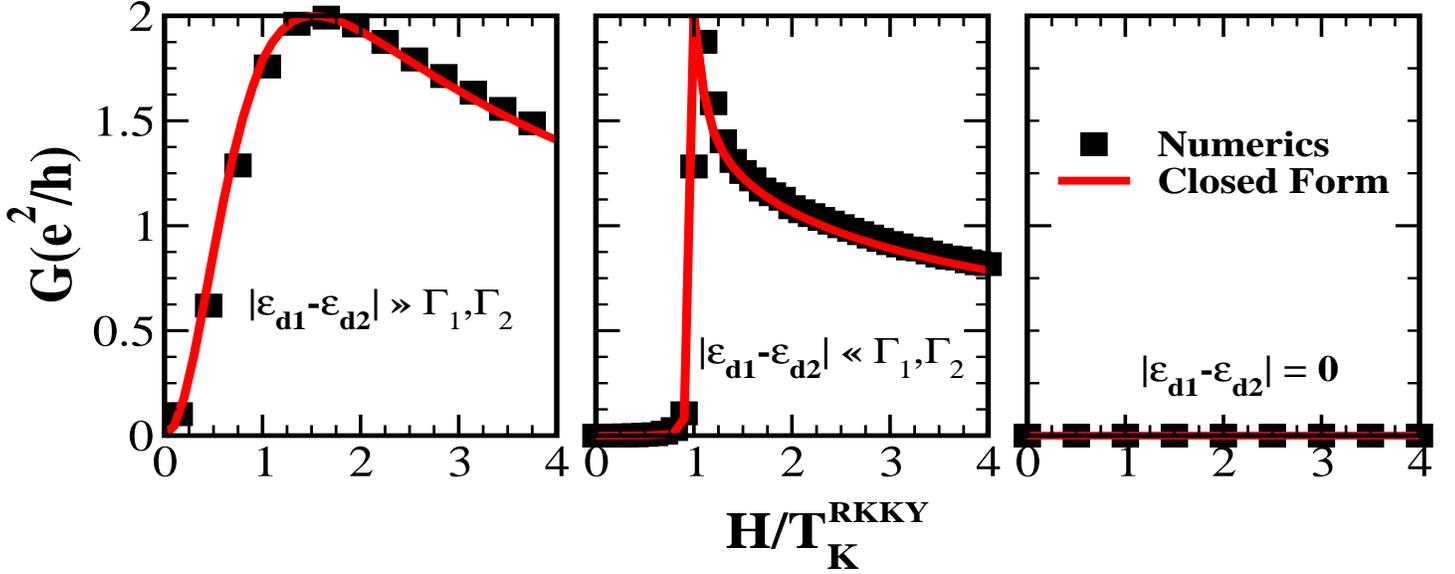}}
\caption{Plotted is the magnetoconductance of the double dot system at its particle-hole symmetric
point.  We consider three cases: right panel ($\delep \gg \Gamma_{1,2}$), middle
panel ($\delep \ll \Gamma_{1,2}$), and left panel ($\delep = 0$).}
\end{figure*}

We now consider the conductance of the double dots at the particle-hole symmetric point.  As we have
discussed in the previous section, this point sees the formation of an RKKY Kondo effect with an
attendant Abrikosov-Suhl resonance.  The presence of this resonance governs intimately how
the conductance evolves as Zeeman fields on the order of the RKKY Kondo
temperature are applied to the dots as it represents the sole available low
energy degrees of freedom.

The shape of the Abrikosov-Suhl resonance thus determines the form of the magnetoconductance
as a function of magnetic field.  In the case that $\delep \gg \Gamma_{1,2}$, the Abrikosov-Suhl
resonance has a Lorentzian form (see the right panel of Figure 4) with a peak centered at
zero energy.  As a Zeeman field is turned on, the magnetoconductance will thus immediately
depart from its particle-hole symmetric value ($0e^2/h$).  The universal form of the $G(H)$
in this case is shown in the rightmost panel of Figure 7.  We plot both the closed form
available from the analysis performed in Section IV.B as well as a direct numerical
evaluation of the Bethe ansatz integral equations.  We see we obtain an excellent match
(thus validating the approximations used in solved these equations analytically).
We see that the magnetoconductance peaks at the maximal possible conductance 
for a value of $H$ corresponding to $1.4 T^{\rm RKKY}_K$.  That is does reach the unitary maximum
is a reflection of the fact that the scattering phases for spin up and down electrons
are symmetric about $2\pi$ (see Eqn. (\ref{eIVxxiii})).  After reaching the unitary maximum
the magnetoconductance decreases uniformly behaving as $1/\log^2(H/T^{\rm RKKY}_K)$
for $H \gg T^{\rm RKKY}_K$.

\begin{figure*}
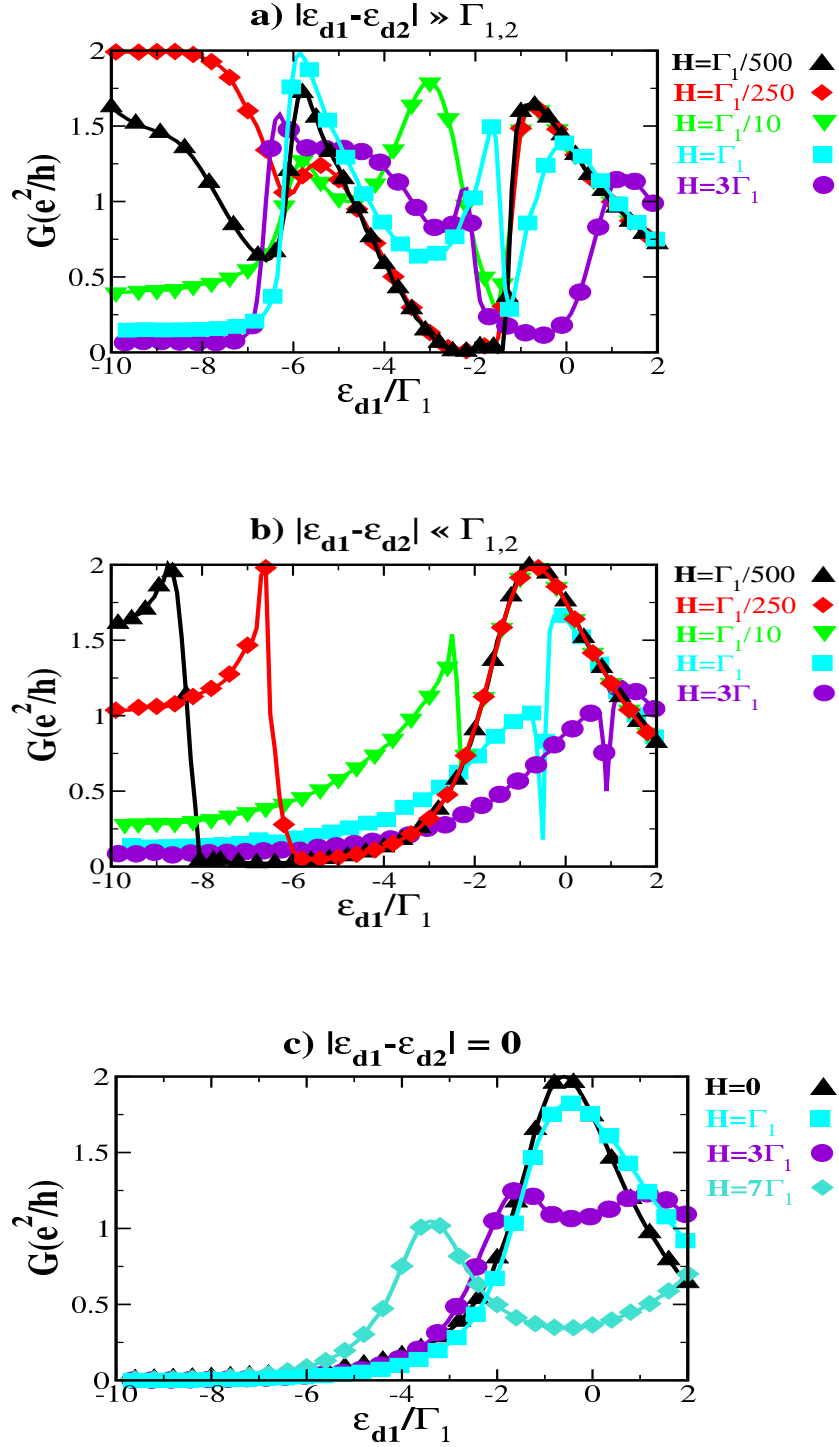

\begin{center}
\epsfysize=1.15\textwidth
\includegraphics[height=5.5cm,width=11cm]{linresH_psep.eps}
\end{center}
\vskip .375in 
\begin{center}
\epsfysize=1.15\textwidth
\includegraphics[height=5.5cm,width=11cm]{linresH_po.eps}
\end{center}
\vskip .375in
\begin{center}
\epsfysize=1.15\textwidth
\includegraphics[height=5.5cm,width=11cm]{linresH_psame.eps}
\end{center}
\vskip .1in
\caption{Plots of the conductance as a function of $\epo$ for a variety of values of Zeeman field, $H$.  We consider
three cases: a) $\delep \gg \Gamma_{1,2}$; b) $\delep \ll \Gamma_{1,2}$; c) $\delep = 0$.  In all three
$U_1 = 1$ and $\Gamma_1 = U/20$.  For a) and b) we see sharp changes in $G$ closely tied to the presence of an
Abrikosov-Suhl resonance at $H=0$.  For c) the resonance is absent and changes in $G$ occur only
when $H$ exceeds $\Gamma_1$.}
\end{figure*}

In the second panel of Figure 7 we plot the magnetoconductance for the case $\delep \ll \Gamma_{1,2}$.
The functional form of the magnetoconductance in this case is also governed by the shape of
the Abrikosov-Suhl resonance. 
As shown in the left panel of Figure 4, the resonance here is peaked at an energy $\epsilon=T^{\rm RKKY}_K$.
While it does not vanish at zero energy, its value relative to its peak value is $4\cos^2(\pi\mu)$
where $\mu$ is defined in Eqn. (\ref{eVii}) and here is approximately 1/2.  Consequently the
change in the conductance in this case upon the initial introduction of a magnetic field will be 
negligible.  But once the magnetic field reaches a strength of $T^{\rm RKKY}_K$, the magnetoconductance
will undergo an extremely rapid variation (as seen in Figure 7).  Once the unitary maximum is
reaches, the magnetoconductance again decreases monotonically behaving as 
$1/\log^2(\htk)$.

We can summarize the behaviour of the magnetoconductance in these two cases via
\begin{widetext}
\begin{eqnarray}\label{eVx}
G = 
\begin{cases}
\frac{2e^2}{h}\frac{\pi^2}{4}\cos^2(\pi\mu)\bigg(\frac{H}{T^{\rm RKKY}_K}\bigg)^2 
+ {\cal O}\bigg(\frac{H}{T^{\rm RKKY}_K}\bigg)^4;  & \text{if $H \ll T^{\rm RKKY}_K$,} \cr
\frac{\pi^2}{2\log^2(\htk)} - \frac{\pi^4\mu^2}{\log^4(\htk)} & \cr
\hskip .5in + {\cal O}\bigg(\frac{\log\log(\htk)}{\log^3(\htk)}\bigg)
+ {\cal O}\bigg(\frac{\mu^2\log\log(\htk)}{\log^5(\htk)}\bigg)
+ {\cal O}\bigg(\frac{\mu^4}{\log^6(\htk)}\bigg),
 & \text{if $H \gg T^{\rm RKKY}_K$,}\cr
\end{cases}
\end{eqnarray}
\end{widetext}
where $\ttk = \frac{2^{3/2}}{\sqrt{\pi e}}\tk$.  In the above expression for the conductance
in the limit $H \gg \ttk$ we have included the leading order term in $\mu^2$ as this governs
the different behaviour in the two cases $\delep \ll \Gamma_{1,2}$ and $\delep \gg \Gamma_{1,2}$.

We now consider the final case when $\epsilon_{d1} = \epsilon_{d2}$.  As discussed previously, in this
instance, the Abrikosov-Suhl resonance identically vanishes.  Without the Abrikosov-Suhl resonance
and the concomitant absence of a low energy density of states,
the conductance is unchanged upon application of a Zeeman field on the order of $\tk$ (see rightmost
panel of Figure 7).  (We note that when $\epsilon_{d1} = \epsilon_{d2}$ there is no Kondo
temperature.  The plot in the righthand panel of Figure 7 is made taking the Kondo temperature
as defined in the case $\delep \sim 0$ but not exactly 0.)

\subsection{Finite Field Linear Response Conductance as a Function of Gate Voltage}

We now consider the magnetoconductance away from the particle-hole symmetric point.
As with this point, the changes induced in the conductance by the application of a magnetic
field is determined by the presence or absence of an Abrikosov-Suhl resonance.  If a resonance
is present, we in general marked changes by the introduction of even a small Zeeman field.
If however the resonance is absent (solely in the case $\delep = 0$), the conductance
will only be sensitive to Zeeman fields on the order of the bare parameters in the model.

\begin{figure}[tbh]
\centerline{\psfig{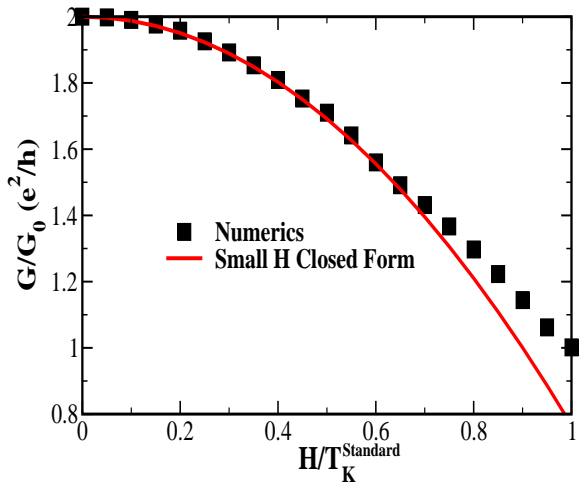}}
\caption{A plot of the magnetoconductance for the case $\delep \gg \Gamma_{1,2}$ at
a value of the gate voltage where one localized electron is found on the dots.  We use the same
parameters as in Figure 8.  We plot
both the results of a numerical analysis of the Bethe ansatz equations as well as the small field
analytic form that would be expected if the system was a single dot.}
\end{figure}

In the top panel of Figure 8 we first consider the case
of $\delep \gg \Gamma_{1,2}$, plotting $G$ as a function of gate voltage (i.e. $\epsilon_{d1}$)
for a variety of values of Zeeman field ranging from $H = \Gamma_1/500$ to $H = 3\Gamma_1$.
As expected from the previous section even the smallest of these field strengths strongly alters
the conductance near the particle hole symmetric point (we note that
$\tk = 0.00243\Gamma_1$).  The region of $\epsilon_{d1}$ over which the conductance is altered from its
zero field value
significantly increases as $H$ is increased.  As $\epsilon_{d1}$ is initially
decreased, we see that the conductance
tracks closely its zero field value (as displayed in the top left panel of Figure 3).  However
a value
of $\epsilon_{d1}$ is inevitably reached where the conductance markedly
changes.  For the smallest field value used, 
$H = \Gamma_1/500$,
this occurs at $\epsilon_{d1} \approx -6\Gamma_1$ while for largest, $H = 3\Gamma_1$, large scale
deviations in $G$ begin at $\epsilon_{d1} \approx \Gamma_1$.

For the smaller values of the field used, the conductance sees a local maximum around 
$\epsilon_{d1} = -6\Gamma_1$.  This is the position of the (standard) Kondo effect we argued to
exist at zero field in the first part of this section.
The Kondo temperature for this effect is $\tks = 0.00422\Gamma_1$, twice
the smallest value of field ($H = .002\Gamma_1$) displayed in Figure 8.  
As the magnetic field is increased, the conductance in the vicinity of $\epsilon_{d1}$
decreases as would be expected for the standard Kondo effect.

We plot this decrease in the conductance maximum as a function of $H$ in Figure 9.
From Ref. (\onlinecite{long}), the expected from of this decrease is
\begin{equation}\label{eVxi}
G(H) = G(H=0)(1 - \frac{\pi^2}{16}\bigg(\frac{H}{\tks}\bigg)^2).
\end{equation}
We see in Figure 9 that the numerical analysis of the Bethe ansatz equations follows this form
for values of the Zeeman field up to $\tks$.

The plot in Figure 9 was produced in the case $\delep \gg \Gamma_{1,2}$.  As this condition
is relaxed, the standard Kondo effect does not disappear.  However the region of field
over which Eqn. (\ref{eVxi}) is valid decreases.

In the middle panel of Figure 8 we consider the magnetoconductance in the case $\delep \ll \Gamma_{1,2}$.
In this case the conductance behaves in much the same fashion as $\delep \gg \Gamma_{1,2}$.  For
the smallest values of the field the conductance goes unaltered from its zero field value
over a substantial range of gate voltages.  This range shrinks as $H$ increases.  And when the conductance
ceases to track its $H=0$ value, it does so abruptly.  Due to the nature of the
Abrikosov-Suhl in this case (i.e. a resonance sharply peaked about $\epsilon = \tk$), we find that the
changes in the linear response conductance, when they occur, to be even more marked.  
Even for fields for in excess of $\tk$ (in this case $\tk = 0.00179\Gamma_1$),
we see sharp changes persist in the magnetoconductance.
For both $H=\Gamma_1$ and $H=3 \Gamma_1$ we see that there are sharp downward changes in the conductance
for values of $\epsilon_{d1}$ near $0$.

In the final panel of Figure 8 we plot the magnetoconductance in the case $\delep = 0$.  Here there
is no Abrikosov-Suhl resonance at the particle-hole symmetric point.  With the absence of a dynamically
enhanced low energy density of states, we find $G$ to be insensitive to values of $H$ smaller than
the bare parameters.  The conductance is approximately equal to its zero field value until $H$ becomes
of the same magnitude as $\Gamma_1$.  For values of $H$ in excess of $\Gamma_1$ we see that the zero
field peak at $\epo \sim -\Gamma_1$) splits in two with increasing separation as $H$ grows.

\section{Finite Temperature Transport of Double Dot}

\begin{figure*}[tbh]
\centerline{\psfig{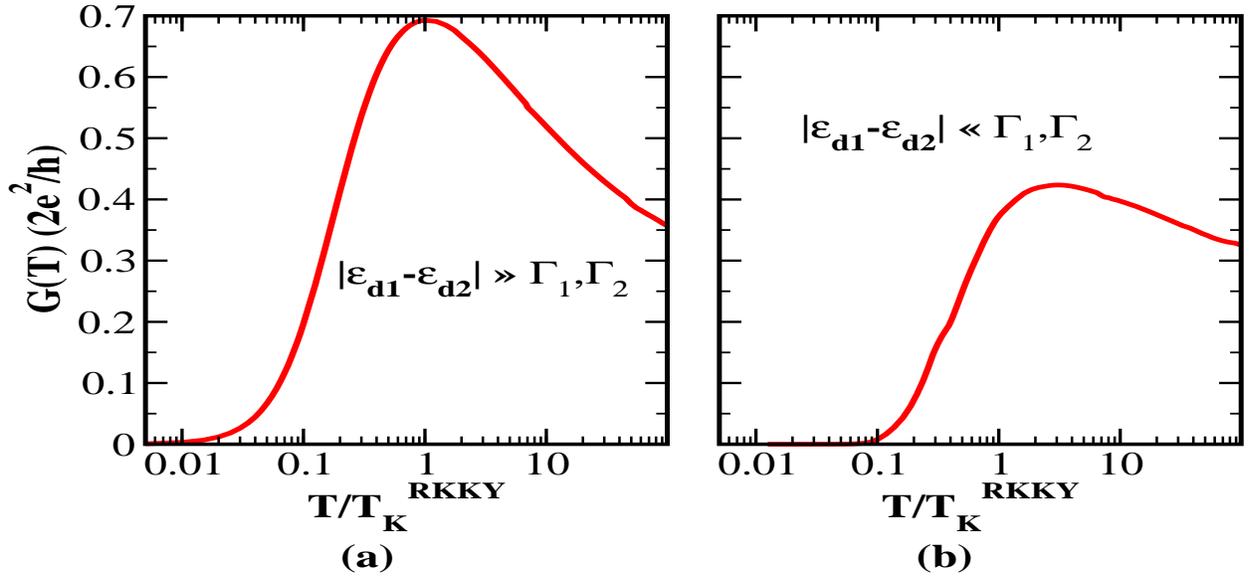}}
\caption{Plots of the finite temperature linear response conductance in the cases
$\delep \gg \Gamma_{1,2}$ (left panel (a)) and $\delep \ll \Gamma_{1,2}$.  In both
cases we employ $U_1 = 1$ and $\Gamma_1 = U/20$.  For (a) we use a bare level
separation of $\ept-\epo = 5\Gamma_1$ while for (b) $\ept-\epo = 0.5\Gamma_1$.}
\end{figure*}

In this section we analyze the finite temperature linear response conductance, $G(T)$, at the particle-hole
symmetric point.  As with the magnetoconductance at this point, the behaviour of $G(T)$ at temperatures
on the order of $\tk$ is ultimately determined by the form of the low energy density of states,
that is, the Abrikosov-Suhl resonance as given in Eqn. (\ref{eViv}).  In particular, the nature
of the Abrikosov-Suhl resonance marks the difference between the two opposing
limits of the bare level separation, i.e. $\delep \gg \Gamma_{1,2}$ and $\delep \ll \Gamma_{1,2}$.

To evaluate the finite temperature conductance, we solve numerically through iteration
the equations in (\ref{eIIIxxi}) and (\ref{eIIIxxv}).  To do so we approximate the source term for the integral
equation governing $\rho^{\rm imp}(k)$ by Eqn. (\ref{eVi}), an excellent approximation provided
we are at the particle-hole symmetric point.  Then following the procedure described in Section
III.C, we obtain the results plotted in Figure 9.

In the case $\delep \gg \Gamma_{1,2}$ (see the left panel of Figure 9), the Abrikosov-Suhl resonance
is peaked at zero energy.  Thus even for temperatures far below the Kondo temperature, $\tk$, we
see an immediate change in the conductance from it zero temperature value ($0e^2/h$).  For
temperatures far below $\tk$ the conductance has a quadratic Fermi liquid form which we can
determine to be 
\begin{equation}\label{eVIi}
G(T) = \frac{2e^2}{h}\frac{\pi^4\cos^2(\mu\pi)}{4}\bigg(\frac{T}{\tk}\bigg)^2.
\end{equation}
We can determine this form by comparing the analysis of the Bethe ansatz equations at
finite temperature for a single level dot.\cite{long}  For a single level dot (sld)
the Abrikosov-Suhl resonance is given by
\begin{equation}\label{eVIii}
\rho^{\rm imp~sld}(\epsilon,T^{\rm sld}_K) = \frac{1}{2T^{\rm sld}_K}\frac{1}{1+\frac{\pi^2\epsilon^2}{4(T^{\rm sld}_K)^2}}.
\end{equation}
For a double dot (dd) with well-separated bare levels (i.e. $\mu \sim 0$), 
the Abrikosov-Suhl resonance has
the same form (see Eqn. (\ref{eViv}))
\begin{equation}\label{eVIiii}
\rho^{\rm imp~dd}(\epsilon,\tk) = 2\cos(\mu\pi)\rho^{\rm imp~sld}(\epsilon,\tk).
\end{equation}
This in turn implies that an electron with energy $\epsilon$ scatters off the double
dot via the relation
\begin{equation}\label{eVIiv}
\delta^{\rm dd}_{\rm el}(T,\epsilon) = 2\cos(\mu\pi) \delta^{\rm sld}_{\rm el}(T,\epsilon).
\end{equation}
This relation follows because the Abrikosov-Suhl resonance serves as the source term
governing the {\it linear} integral equations determining the scattering phase.
For a single level dot, the deviation of the 
finite temperature linear response conductance from its zero temperature value takes the form,\cite{long,costi}
$$
\delta G(T) = \frac{2e^2}{h}\frac{\pi^4}{16}\frac{T^2}{(T^{\rm sld}_K)^2}.
$$
As $G(T) = \int \partial_E f(E) \sin^2(\delta_{\rm el}(E,T))$, we see that for $\delta_{\rm el}(E)$ small,
the low temperature form of $G(T)$ for the single level dot determines $G(T)$
for the double dot to be given as in Eqn. (\ref{eVIi}).

As the temperature is increased, $G(T)$ reaches a maximum at roughly $T = \tk$.  With further
increases, $G(T)$ decreases from its maximal value (for the parameters chosen, $1.4 e^2/h$)
gradually.  For large values of temperature we see
that the conductance decreases logarithmically.  From the behaviour of $G(T \gg T_K)$ for a single level dot 
(Ref. (\onlinecite{kaminski}))
and the similarity of analysis between that case and a double dot, we might expect
\begin{equation}\label{eVIv}
G(T\gg \tk ) \sim \frac{1}{\log^2(T/\tk)}.
\end{equation}
However from examining panel (a) of Figure 9, it is clear that if this behaviour does hold,
it is valid only for temperatures much larger than $10\tk$ (in the range $10\tk - 100\tk$,
Figure 9 implies the conductance is decreasing at best according to $1/\log(T/\tk)$).  This however is
unsurprising as a similar condition holds on the finite temperature behaviour of a single
level dot for $T \gg T_K$.

In panel (b) of Figure 9 we plot $G(T)$ for the case of $\delep \ll \Gamma_{1,2}$.  In this case
the Abrikosov-Suhl resonance has a sharp peak at an energy $\ep = \tk$ with minimal (although
finite) spectral weight at zero energy.  We thus expect that the low temperature response
to be much smaller than for the case $\delep \gg \Gamma_{1,2}$.  And we see this
in Figure 9 where $G(T)$ is effectively $0e^2/h$ (equal to its zero temperature value)
until $T \sim 0.1 \tk$.  For $T \ll \tk$, $G(T)$ has the exact same form as given in Eqn. (\ref{eVIi})
for the case $\delep \gg \Gamma_{1,2}$.  However here $\mu \sim 1/2$ and so the coefficient
of the $T^2$ correction is extremely small.

As the temperature is increased in this case to roughly $2\tk$, a maximum occurs in the conductance
(for the particular parameters chosen).  While the value at which the maximum occurs
is non-universal, we do expect it to roughly occur when the temperature is ${\cal O}(\tk)$.
With further increases in temperature, the conductance logarithmically decreases.  
If this decrease is to behave as $G \sim 1/\log^2(T/\tk)$ (as for a single level dot),
the temperature at which this asymptotic form begins to be valid is far in excess of $10\tk$.

\section{Discussion}

\subsection{Friedel Sum Rule}

The Friedel sum rule relates the scattering phase, $\delta_{\sigma e}$, of an electron of spin $\sigma$
at the Fermi surface at zero temperature
to the number of electrons displaced by connecting the leads to the dot, $n_{\sigma\rm dis}$:
\begin{equation}\label{eVIIi}
\delta_{\sigma e} = 2\pi n_{\sigma\rm dis}.
\end{equation}
$n_{\sigma \rm dis}$ has two contributions: {\bf 1)} the number of localized electrons sitting
on the dot and equal to,
$$
\sum_{\alpha}\langle d^\dagger_{\sigma\alpha}d_{\sigma\alpha}\rangle
$$
and {\bf 2)} the $1/L$ deviations induced in the electron number in the leads
$$
\delta n_{\rm leads} = \int dx \bigg[\langle c^\dagger_{\sigma e}(x)c_{\sigma e}(x) - \rho_{\sigma \rm bulk}\bigg],
$$
where $\rho_{\sigma \rm bulk}=L/2\pi$ is the bulk interacting density of states
of the leads with the leads and dots unconnected to one another.  While this second
contribution is typically not considered in the application of the Friedel sum rule,
its proof explicitly takes it into account.\cite{langreth}

We have already seen that in the model of double dots analyzed here that this second contribution, 
$\delta n_{\rm leads}$,
is non-negligible and in fact can be negative, driving $n_{\sigma\rm dis}$ itself negative (and thus
cannot be ignored if the conductance is to be computed correctly).
The range of gate voltages where this occurs marks a region of rapid variation in the conductance
(see the top left panel of Figure 3 for example).  The possibility that this second contribution
is non-zero is closely tied to the presence of two interfering pathways for the electrons
to transit the dot system.  A finite value for $\delta n_{\rm leads}$ has also been seen in the analysis
of Fano resonances in quantum dots.\cite{fano}  For dots exhibiting Fano resonances, there are 
two tunneling paths between the leads: one resonant
inducing a single particle scattering phase, $\delta (q) = -2\tan^{-1}(\Gamma/(q-\epsilon_d))$
and a non-resonant second path
with a constant scattering phase.  The two paths, however, interfere and the overall scattering phase
is more complicated than the mere sum of the two (akin to the fact the sum of the
bare scattering phase off the double dots in Eqn. (\ref{eIIv}) has the sum in the argument of 
$\tan^{-1}$).  The system exhibiting Fano resonances
may be considered a certain limit of the double dots
studied here.  If we take the limit of $\epsilon_{d2}$ and $\Gamma_2$ in such a way that
$$
\frac{\Gamma_2}{q-\epsilon_{d2}}
$$
is constant for all wavevectors $q$ we recover the physics of Fano resonances.

$n_{\sigma\rm dis}$ is a quantity naturally available from the Bethe ansatz as the Bethe ansatz
allows one to directly access the scattering phase, i.e. the formalism, as discussed in Section III,
provides a means to derive the scattering
phase independent of directly invoking the Friedel sum rule.  In this sense
we provide a proof of the Friedel sum rule in the case of a double dot system.
We do note that however we are able to compute 
in principle the occupation of the dots $\sum_{\alpha}\langle d^\dagger_{\sigma\alpha}d_{\sigma\alpha}\rangle$
by computing the impurity free energy and then taking appropriate derivatives with respect to $\epsilon_{d\alpha}$.

\subsection{The Effects of Breaking Integrability}

We now turn to the question of how fined tuned the results presented herein are.  Our ability to
exactly solve the models of the double dots depends upon the constraints listed in Eqn. (\ref{eIIix}) being
satisfied.  However it is important to determine how the physics is affected if these conditions are
broken, at least weakly.  While it is easily conceivably that an experimental double dot system
could be tuned so that its parameters satisfy approximately the constraints of Eqn. (\ref{eIIix}), we
cannot hope that the constraints are met {\it exactly}.

Fortuitously, we do not expect any of the major phenomena identified in this paper to change qualitatively 
if Eqn. (\ref{eIIix})
is weakly violated.
Generally, we expect the physics to be perturbative in any deviation 
from the conditions necessary for integrability.  
To see this, contrast the situation of a single dot coupled to a single lead.  This problem is non-perturbative
in the dot-lead coupling, a partial consequence of the ambiguous nature of the zeroth order
zero temperature dot density matrix about which one conducts the perturbation theory.  This is true
both in the presence and absence of interactions.  Indeed it is relatively straightforward
to derive the dots' Green functions at $U=0$ and see that they do not have a well-defined $V \rightarrow 0$
limit.
Concomitantly, perturbation
theory in $U$, the dot Coulomb repulsion, is well behaved as the dot Green's function is unambiguously
defined at $U=0$ by the presence of the dot-lead hybridization.
Similarly if we weakly break one of the integrability constraints, we expect the deviation
to behave in a perturbative fashion as the dot state is already robustly established.
This principle guided the study of perturbations about the exactly solvable Toulouse
limit of a non-equilibrium dot in Ref. (\onlinecite{hershfield}).

Now while small deviations from the integrability constraints will only quantitatively change the physics,
the same cannot be said of large deviations.  In Ref. (\onlinecite{dias}) a particle-hole symmetric double dot is considered
where $\epo = - U_1/2$, $\ept = U_2=0$, and $V_1,V_2$ finite.  These conditions represent a strong violation
of the integrability condition, $V^2_2U_2 = V^2_1U_1$.  And instead of a Kondo effect, the double dot
is found to be in a local moment regime.

There are other more specific arguments as to why violating the integrability
constraints will not drastically change the physics.
To this end, let us consider the constraint upon the left-right hopping amplitudes.  If this constraint
is met, only one electron channel couples to the dot (the even channel), while the second
channel is decoupled.  Let us consider the affects of weakly violating this condition.
The second (odd) channel then couples to the dots.  However this will change qualitatively 
none of the 
Kondo physics discussed in the manuscript.  
In principle, the second channel {\it could} introduce a second Kondo
scale, $T_{K2}$ 
(for either the RKKY Kondo effect or the regular Kondo effect) as in 
Ref. (\onlinecite{pustilnik}).  If we start at some high temperature and begin to lower it, 
when $T$ reaches $T_{K1}$ (the Kondo temperature associated with the even channel
and given in the manuscript) the Kondo singlets (either RKKY or standard) will
establish themselves.  If we continue to lower the temperature to $T_{K2}$ nothing further will happen
as there are no spin degrees of freedom (the state is a singlet) left to screen.  
(This contrasts with Ref. (\onlinecite{pustilnik}) where an effective spin-S ($S\geq 1$) impurity was
envisioned and a two-stage Kondo effect was predicted to occur).

Another argument that demonstrates the RKKY Kondo effect is not a result of fine tuning 
makes appeal to a Schrieffer-Wolfe transformation at the particle-hole symmetric point.
Under such a transformation the Hamiltonian reduces to the 
form,
\begin{equation}\label{eVII}
H = H_{\rm conduction electrons} + J_1 S_1 \psi^\dagger\sigma\psi + J_2 S_2 \psi^\dagger\sigma\psi,
\end{equation}
where the two couplings, $J_i$, are given by 
$$
J_i = \frac{2\Gamma_i}{U+\epsilon_{di}} - \frac{2\Gamma_i}{\epsilon_{di}}.
$$
We see that generically $J_1\neq J_2$ for the integrable range of parameters 
and that deviations away from the integrability
constraints do not change this fact.  Thus we would 
expect such 
deviations to affect only non-universal aspects of the physics (such as the exact values of Kondo temperatures).

While a small breaking of the integrability conditions will not destroy the RKKY Kondo effect, there is
one small perturbation that will.  The RKKY Kondo effect involves singlet formation of two electrons
sitting on two dots as mediated by the electrons sitting in the leads.  But if one were to add a coupling,
$J_{\rm singlet}$, in excess of the Kondo temperature, $\tk$, 
that promoted direct singlet formation between the two localized electrons, i.e $\delta H = J_{\rm singlet}
S_1\cdot S_2$, we would expect to destroy the RKKY Kondo effect.  In principle, any experimental
realization of a double dot system will contain such a $J_{\rm singlet}$.  And because
$\tk$ is exponentially small compared to the bare parameters, one might think that the RKKY
Kondo effect will be generically overcome.  While this may be true at the particle-hole symmetric
point, the RKKY Kondo effect exists over a range of gate voltages centered around this point.  By
tuning the gate voltage to different values within this range, one can increase $\tk$ above any
putative value of $J_{\rm singlet}$.

We finally turn to the effects of breaking the integrability conditions on the 
the quantum critical point at $\epsilon_{d1}=\epsilon_{d2}$.  We note 
that this point is tied to the disappearance of 
a singular feature in the non-interacting impurity density of states at $\epsilon_{d1}=\epsilon_{d2}$
(see Figure 6).  In turn this disappearance is dependent upon the decoupling of the odd dot
degree of freedom, $d_{odd}$.  A perturbation that restores this coupling 
(such as weakly coupling a second channel of electrons to the dots)
will change the critical point
to a smooth crossover and so restore the RKKY Kondo effect,
but a perturbation which leaves $d_{odd}$ uncoupled (such as perturbations on the inter-/intra-dot Coulomb repulsion)
will presumably leave the critical point in place.

\acknowledgments
This work was supported by the DOE under contract DE-AC02-98 CH 10886.

\appendix

\section{Wiener-Hopf Analysis of Equations Governing the Linear Response Conductance}

\subsection{Review of Wiener-Hopf Analysis}

We review the method of solution as presented in Ref. (\onlinecite{wie})
for equations of the general form
\begin{equation}\label{eAi}
f(z) = \int^A_{-\infty} dz' f(z')h(z-z') + g(z).
\end{equation}
Writing $f^\pm (z) = f(z)\theta (\pm z \mp A)$, the Fourier transform
of the above equation yields
\begin{equation}\label{eAii}
f^+(\om ) + f^-(\om ) = f^- (\om ) h(\om ) + g(\om ),
\end{equation}
where Fourier transforms are defined by
\begin{equation}\label{eAiii}
a (\om ) = \int d\om e^{i\om z} a (z).
\end{equation}
The key step in the analysis is writing 
$1-h(\om)$ as a product of functions, $G^\pm$, which are analytic in the
upper/lower planes respectively:
\begin{equation}\label{eAiv}
1 - h(\om ) = \frac{1}{G^+(\om ) G^- (\om )}.
\end{equation}
We can then write Eqn.(\ref{eAii}) as
\begin{equation}\label{eAv}
e^{-i\om A} \frac{f^-(\om )}{G^-(\om )} + 
e^{-i\om A} f^+(\om ) G^+(\om )  =  g(\om )G^+(\om ) e^{-i\om A}.
\end{equation}
Given $e^{-i\om A}f^\pm (\om)$ is analytic in the upper/lower half plane,
applying the operators
\begin{equation}\label{eAvi}
\pm \frac{1}{2\pi i} \int d\om ' \frac{1}{\om ' - (\om \pm i\delta)},
\end{equation}
to Eqn.(\ref{eAv}) yields solutions for both $f^+$ and $f^-$:
\begin{eqnarray}\label{eAvii}
f^-(\om ) &=& -G^-(\om ) \frac{e^{i\om A}}{2\pi i} \cr
&& \hskip -.4in \times \int d\om ' 
\frac{1}{\om ' - (\om - i\delta)} g(\om ') G^+(\om ')e^{-i\om 'A};\cr\cr
f^+(\om ) &=& \frac{e^{i\om A}}{G^+(\om )} \frac{1}{2\pi i} \cr
&& \hskip -.4in \times \int d\om ' 
\frac{1}{\om ' - (\om + i\delta)} g(\om ') G^+(\om ')e^{-i\om 'A}.
\end{eqnarray}

\subsection{A Wiener-Hopf Analysis of the Scattering Phase Analysis of Zero Field}
The scattering phase can be written as
\begin{eqnarray}\label{eAviii}
\delta_e &=& 2\pi\int^\infty_Q d\lambda (\sigma_1(\lambda)+\sigma_2(\lambda));\cr\cr
&=& 2\pi - \pi\int^Q_{-\infty} d\lambda (\sigma_1(\lambda)+\sigma_2(\lambda)).
\end{eqnarray}
where each $\sigma_i(\lambda)$ satisfies
\begin{eqnarray}\label{eAvix}
\sigma^{\rm imp}_i (\lambda) &=& \int^Q_{-\infty}d\lambda' R(\lambda-\lambda')\sigma^{\rm imp}_i(\lambda ')\cr
&& \hskip -.6in +{\rm sgn}(1-\beta_i)\int^\infty_{-\infty}d\lambda'(1+a_2)^{-1}(\lambda-\lambda')a_{|1-\beta_i^{-1}|}(\lambda')\cr
&& \hskip -.6in +\int^\infty_{-\infty}dq s(\lambda - g(q))\Delta_{0i}(q).
\end{eqnarray}
The above two equations can be obtained by inverting via Fourier transform Eqn.(\ref{eIVviii}).

Identifying 
\begin{eqnarray}\label{eAx}
g(\omega ) &=& \frac{1}{1+e^{-|\omega|}}\bigg(e^{-|\omega|/2}
\int^\infty_{-\infty}dq \Delta_{0i}(q)e^{i\omega g(q)} \cr\cr
&& + {\rm sign}(1-\beta_i)\frac{e^{-\gamma_i|\omega|+i\omega I_i^{-1}}}{1+e^{-|\omega|}}\bigg);\cr\cr
\gamma_i &=& -\frac{1}{2}(1-\beta_i^{-1}){\rm sign}(1-\beta_i);\cr\cr
I_i^{-1} &=& \frac{\alpha_i^2-\tilde\Gamma_i^2}{2U_i\Gamma_i};\cr\cr
G^\pm(\omega ) &=& \frac{\sqrt{2\pi}}{\Gamma(1/2\mp i\omega/2\pi)}
\bigg(\frac{\mp i\omega + \delta}{2\pi e}\bigg)^{\mp i\omega/2\pi},
\end{eqnarray}
where $\beta_i$ and $\alpha_i$ are defined in Eqns.(\ref{eIVv}),
we apply the Wiener-Hopf methodology presented at the beginning of this section
to compute
\begin{equation}\label{eAxi}
\int^Q_{-\infty} d\lambda \sigma_i(\lambda) = \sigma_i^-(\omega = 0).
\end{equation}
We so find
\begin{eqnarray}\label{eAxii}
\int^Q_{-\infty} d\lambda \sigma_i(\lambda) = K_{i1} + K_{i2};
\end{eqnarray}
where
\begin{eqnarray}\label{eAxiii}
K_{i1} &=& -\frac{1}{2\pi^{3/2} i}\int d\omega \frac{1}{\omega+i\delta}
\big(\frac{-i\omega+\epsilon}{2\pi e}\big)^{-\frac{i\omega}{2\pi}}\cr\cr
&& \hskip -.5in \times \Gamma(\frac{1}{2}+\frac{i\omega}{2\pi})
\int^\infty_{-\infty}dq \Delta_{0i}(q)e^{i\omega (g(q)-Q)} \cr\cr
K_{i2} &=&  {\rm sign}(\beta_i-1)\frac{1}{2\pi^{3/2} i}\int d\omega \frac{1}{\omega+i\delta}
(-i\omega+\delta)^{-\frac{i\gamma_i\omega}{2\pi}}\cr\cr
&& \hskip -.5in \times (i\omega+\delta)^{\frac{i(\gamma_i-1/2)\omega}{\pi}}
\Gamma(\frac{1}{2}+\frac{i\omega}{2\pi})
(\frac{1}{2\pi e})^{-\frac{i\omega}{2\pi}}e^{-i\omega (Q-I_i^{-1})}.\cr
&&
\end{eqnarray}
We will evaluate each term $K_{i1/2}$ in turn.

For $K_{i1}$, if $Q<0$, we can continue the contour into the upper half plane with
the result
\begin{eqnarray}\label{eAxiv}
K_{i1}(Q<0) &=& \frac{1}{\pi}\sum^\infty_{n=0}\frac{(-1)^n}{(n+1/2)\Gamma(n+1)}
(\frac{n+1/2}{e})^{n+1/2}\cr\cr
&& \hskip -.5in \times \int^\infty_{-\infty} dq \Delta_{0i}(q) e^{-\pi (2n+1)(g(q)-Q)}.
\end{eqnarray}
If instead $Q>0$, we directly evaluate the integral $\int dq$,
\begin{eqnarray}\label{eAxv}
\int^\infty_{-\infty} dq \Delta_{0i}(q) e^{i\omega g(q)} 
\approx e^{i\omega I_i^{-1}-\frac{\tilde\Gamma_i\alpha_i}{U_i\Gamma_i}|\omega|}.
\end{eqnarray}
Setting $J_i=I_i^{-1}-Q$ we can then write $K_{i1}$ as
\begin{eqnarray}\label{eAxvi}
K_{i1} &=& \frac{i}{\sqrt{2}\pi}\int^\infty_{-\infty}dw 
\frac{e^{i\omega J_i-|\frac{\omega}{2}|(1+\beta_i^{-1})}}{G^-(\omega)(\omega+i\epsilon)}.
\end{eqnarray}
If $J_i<0$, we evaluate this integral by continuing into the lower half plane:
\begin{eqnarray}\label{eAxvii}
K_{i1}(Q>0,J_i<0) \!\!&=&\!\! 1 - \frac{1}{\pi^{3/2}}\int^\infty_0 d\omega
\Gamma(\frac{1}{2}+\omega)\cr\cr
&& \hskip -.7in \times 
\frac{\sin(\pi\omega(1+\beta_i^{-1})}{\omega}
(\frac{e}{\omega})^\omega e^{2\pi J_i\omega}\cr\cr
\!\! 
&=& \!\! 1 - \frac{1+\beta_i^{-1}}{2\pi J_i} + {\cal O}(\frac{1}{J_i^2}),
\end{eqnarray}
where in the last line we have written down an expansion in $J_i$ to leading
order in $1/J_i$.
If instead $J_i>0$, i.e. $0<Q<I_i^{-1}$,
we close the contour in the upper-half plane obtaining,
\begin{eqnarray}\label{eAxviii}
K_{i1} &=& \frac{1}{\sqrt{\pi}}{\bf P}\int^\infty_0 \frac{d\omega}{\omega}
\frac{\sin(\pi\beta_i^{-1}\omega)}{\cos(\pi\omega)}\bigg(\frac{\omega}{e}\bigg)
\frac{e^{-2\pi J_i\omega}}{\Gamma(\frac{1}{2}+\omega)}\cr\cr
&& +\frac{1}{\sqrt{\pi}}\sum^\infty_{n=0}\frac{(-1)^n}{n+\frac{1}{2}}\bigg(\frac{n+\frac{1}{2}}{e}\bigg)^{n+\frac{1}{2}}\cr\cr
&& \hskip .35in 
\times \frac{\cos(\pi\beta_i^{-1}(n+\frac{1}{2}))e^{-2\pi J_i(n+\frac{1}{2})}}{\Gamma(1+n)},
\end{eqnarray}
where the symbol ${\bf P}$ indicates that the principal value of the integral
is to be taken.
We can alternatively 
evaluate this integral and sum in a fashion similar to that used in Refs. (\onlinecite{wie})
and (\onlinecite{long})
to obtain,
\begin{eqnarray}\label{eAxix}
K_{i1}(Q>0,J_i>0) &=& \sqrt{2} - 1 - \frac{\pi}{\sqrt{2}6}J_i + \frac{\pi^2\sqrt{2}}{24^2}J_i^2 \cr\cr
&& \hskip -1.45in +\frac{1}{\sqrt{2}\pi}(\beta_i^{-1}-1)J_i 
+ {\cal O}((\beta_i^{-1}-1)^2J_i) + {\cal O}(J_i^3).
\end{eqnarray}

We similarly evaluate $K_{i2}$ for $J_i>0$ and $J_i<0$.  For $J_i>0$ we can continue
into the upper half plane with the result
\begin{eqnarray}\label{eAxx}
K_{i2}(J_i>0) &=& \frac{{\rm sign}(1-\beta_i)}{\sqrt{\pi}}{\bf P}
\int^\infty_0 d\omega (\frac{\omega}{e})^\omega
e^{2\pi J_i\omega}\cr\cr
&& \hskip -.7in \times \frac{1}{\Gamma(\frac{1}{2}+\omega)\cos (\pi\omega)}
\frac{\sin(2\pi\omega(\gamma_i-\frac{1}{2}))}{\omega}\cr\cr 
&& \hskip -.7in + \frac{{\rm sign}(1-\beta_i)}{\sqrt{\pi}}\sum^\infty_{n=0}\frac{(-1)^n}{n+\frac{1}{2}}
\bigg(\frac{n+\frac{1}{2}}{e}\bigg)^{n+\frac{1}{2}}\cr\cr
&& \hskip -.2in \times \frac{e^{-2\pi J_i\omega}\cos(\pi(2n+1)(\gamma_i-\frac{1}{2})}{\Gamma(1+n)}\cr\cr
&& \hskip -.5in = \frac{{\rm sign}(1-\beta_i)(\gamma_i-1/2)}{\pi J_i} + {\cal O}(\frac{1}{J_i^2}).
\end{eqnarray}
And if $J_i<0$, we instead continue into the lower half plane with the result,
\begin{eqnarray}\label{eAxxi}
K_{i2}(J_i<0) &=& {\rm sign}(1-\beta_i)\bigg(
1 - \frac{1}{\pi^{3/2}}
\int^\infty_0 d\omega
e^{2\pi J_i\omega}\cr\cr
&& \hskip -.7in \times (\frac{w}{\omega})^\omega \Gamma(\frac{1}{2}+\omega)
\frac{\sin(2\pi\omega\gamma_i)}{\omega}\bigg)\cr\cr
&&\hskip -.5in ={\rm sign}(1-\beta_i)\big(1-\frac{\gamma_i}{\pi |J_i|}\big)+ {\cal O}(\frac{1}{J_i^2})\big).
\end{eqnarray}

We now finally turn to how $Q$ depends on $U_i$, $\Gamma_i$ and $\epsilon_{di}$.
As the bulk density equations are the same as for a single dot, this dependence
can be taken directly from Refs. (\onlinecite{wie}) and (\onlinecite{long}).  If $Q<0$, $Q$ is determined
implicitly by the equation
\begin{eqnarray}\label{eAxxii}
\frac{2\epsilon_{di}+U_i}{\sqrt{2U_i\Gamma_i}}\!\!\! &=&\!\!\!
\frac{\sqrt{2}}{\pi}\sum^\infty_{n=0}
\frac{(-1)^ne^{\pi Q(2n+1)}}{(2n+1)^{3/2}}G^+(i\pi(2n+1)).\cr
&&
\end{eqnarray}
If further $Q\gg 0$, $Q$ is given by
\begin{eqnarray}\label{eAxxiii}
Q &=& q^* + \frac{1}{2\pi}\ln (2\pi e q^*)\cr\cr
\sqrt{q^*} &=& \frac{\epsilon_{di}+U_i/2}{\sqrt{2U_i\Gamma_i}}.
\end{eqnarray}
We note that there are no inconsistencies in the above equations 
as the the quantities $2U_i\Gamma_i$ and $U_i + 2\epsilon_{di}$ are independent of $i=1,2$.

\subsection{Analysis of Finite Field Equations} 

The equation we must solve takes the form
\begin{eqnarray}\label{eAxxiv}
\rho^{\rm imp}_i (q) &=& \rho^{\rm imp}_{0i}(q) \cr
&& \hskip -.5in - g'(q)\int^B_{-\infty}d\lambda' 
R(g(q') - g(q))\rho^{\rm imp}_{i}(q').
\end{eqnarray}
As we are interested in the behaviour of $\rho^{\rm imp}_i(q)$ at
$q<0$, we introduce the change of variables $z=g(q)$ for $z>0$ and $q<0$
and define $\rho^{\rm imp}_i(z) = -\rho^{\rm imp}_i(q)/g'(q)$.
The above equation is then recast as
\begin{eqnarray}\label{eAxxv}
\rho^{\rm imp}_i (z) \!\!&=&\!\! \rho^{\rm imp}_{0i}(z) \!\!+\!\! \int^\infty_b dz R(z-z')\rho^{\rm imp}(z'),
\end{eqnarray}
where $b = B^2/(2U_i\Gamma_i)$ and 
we can write the Fourier transform of $\rho^{\rm imp}_{0i}(z)$ as
\begin{eqnarray}\label{eAxxvi}
\rho^{\rm imp}_{0i}(\omega ) &=& \int d\omega e^{i\omega z}\rho^{\rm imp}_{0i}(z )\cr\cr
&=& \Theta(\beta_i-1)\frac{\cosh(\omega(\frac{1}{2}-\frac{1}{2\beta_i}))e^{i\omega I_i^{-1}}}
{\cosh(\frac{\omega}{2})} \cr\cr
&& + \frac{i}{2}\tanh\frac{\omega}{2}\int dq e^{i\omega g(q)}\Delta_{0i}(q).
\end{eqnarray}
The kernel, $R(z)$, of the integral equation in Eqn. (\ref{eAxxv}) is the same encountered in analyzing
the scattering phase at zero field.  Thus using the analysis at the beginning of this Appendix
together with the properties of $R(z)$ already established, 
we can immediately write down the form of $\rho^{\rm imp}_i (z)$:
\begin{eqnarray}\label{eAxxvii}
\int^B_{-\infty} dq \rho^{\rm imp}_i(q) &=& \frac{G_+(0)}{2\pi i}\int^\infty_{-\infty}d\omega 
\frac{e^{-i\omega b}\rho^{\rm imp}_{0i}(\omega)G_-(\omega)}{\omega - i\delta}.\cr
&&
\end{eqnarray}
Substituting in the above form for $\rho^{\rm imp}_{0i}(\omega )$, we obtain the
solution as written in Eqn. (\ref{eIVxxvii}).  As discussed in Section V.B, the Fermi surface parameter, $b$,
is the same as found in the analysis of the Anderson model for single level dots and
for small fields takes the form $b = -\frac{1}{2\pi}\log\frac{\pi e H^2}{4U_i\Gamma_i}$.



\begin{thebibliography}{99}





\bibitem{gold} D. Goldhaber-Gordon et al.
PRL {\bf 81}, 5225 (1998); D. Goldhaber-Gordon et al.,
Nature {\bf 391}, 156 (1998).

\bibitem{kondo}
S. Cronenwett et al., Science 281, {\bf 540}, (1998);
W.G. van der Wiel et al., Science {\bf 289}, 2105 (2000).

\bibitem{eto}
M. Eto and Y. Nazarov, Phys. Rev. Lett. {\bf 85}, 1306 (2000).

\bibitem{pus}
 M. Pustilnik, L. I. Glazman, Phys. Rev. Lett. {\bf 87}, 216601 (2001).

\bibitem{pus1}
 M. Pustilnik, L. I. Glazman, Phys. Rev. Lett. {\bf 85}, 2993 (2000).

\bibitem{wiel}
W.G. van der Wiel, S. De Franceschi, J.M. Elzerman, S. Tarucha, L.P. Kouwenhoven, J. Motohisa, F. Nakajima, T. Fukui,
Phys. Rev. Lett. {\bf 88}, 126803 (2002).

\bibitem{RKKY1} 
H. Jeong, A. M. Chang, and M. R. Melloch, Science {\bf 293}, 2221 (2001). 

\bibitem{RKKY2} 
N. J. Craig, J. M. Taylor, E. A. Lester, C. M. Marcus, M. P. Hanson, and A. C. Gossard, Science {\bf 304}, 565 (2004). 

\bibitem{scdots1}
J. R. Petta, A. C. Johnson, J. M. Taylor, E. A. Laird, A. Yacoby, M. D. Lukin, C. M. Marcus, M. P. Hanson, 
and A. C. Gossard, Science {\bf 309}, 2180 (2005).

\bibitem{scdots2} 
F. H. L. Koppens, J. A. Folk, J. M. Elzerman, R. Hanson, L. H. Willems van Beveren, 
I. T. Vink, H. P. Tranitz, W. Wegscheider, L. P. Kouwenhoven, and L. M. K. Vandersypen, Science {\bf 309}, 1346 (2005).

\bibitem{cndots1}
N. Mason, M. J. Biercuk, and C. M. Marcus, Science {\bf 303}, 655 (2004). 

\bibitem{cndots2} 
Sami Sapmaz, Carola Meyer, Piotr Beliczynski, Pablo Jarillo-Herrero, and Leo P. Kouwenhoven, 
Nano Lett. {\bf 6} 1350 (2006).

\bibitem{dec1} A. C. Johnson, J. R. Petta, J. M. Taylor, A. Yacoby, M. D. Lukin, C. M. Marcus, M. P. Hanson, A. C. Gossard,
Nature  {\bf 435} 925 (2005).

\bibitem{dec3} J. Taylor, J. Petta, A. Johnson, A. Yacoby, C. Marcus, and M. Lukin, cond-mat/0602470.

\bibitem{dec4} A. S. Bracker, E. A. Stinaff, D. Gammon, M. E. Ware, J. G. Tischler, A. Shabaev, Al. L. Efros, D. Park,
D. Gershoni, V. L. Korenev, and I. A. Merkulov, Phys. Rev. Lett. {\bf 94}, 047402 (2005).

\bibitem{simon2} R. Lopez, D. Sanchez, M. Lee, M.-S. Choi, P. Simon, and K. Le Hur, Phys. Rev. B 71, 115312 (2005).

\bibitem{c4} 
R. Lopez, R. Aguado, and G. Platero, Phys. Rev. Lett. {\bf 89}, 136802 (2002).

\bibitem{c1} Y. Tanaka and N. Kawakami,
Phys. Rev. B {\bf 72}, 085304 (2005).

\bibitem{c5} E. Vernek, N. Sandler, S. E. Ulloa, E. V. Anda,
cond-mat/0511226.

\bibitem{c7} C.A. B\"usser, G.B. Martins, K.A. Al-Hassanieh, Adriana Moreo, and Elbio Dagotto,
Phys. Rev. B {\bf 70}, 245303 (2004).

\bibitem{c9} D. Boese, W. Hofstetter, and H. Schoeller,
Phys. Rev. B {\bf 66}, 125315 (2002).

\bibitem{schiller} V. Kashcheyevs, A. Schiller, A. Aharony, O. Entin-Wohlman, cond-mat/0610194.

\bibitem{glazman} M. Vavilov, L. Glazman, Phys. Rev. Lett {\bf 94}, 086805 (2005).

\bibitem{grempel} P. S. Cornaglia and D. R. Grempel, Phys. Rev. b {\bf 71}, 075305 (2005).

\bibitem{simon} P. Simon, R. Lopez, and Y. Oreg, Phys. Rev. Lett. {\bf 94}, 086602 (2005).

\bibitem{georges} A. Georges and Y. Meir, Phys. Rev. Lett. {\bf 82}, 3508 (1999).

\bibitem{pos} A. Posazhennikova and P. Coleman, Phys. Rev. Lett. {\bf 94}, 036802 (2005).

\bibitem{dias} L. Dias da Silva, N. P. Sandler, K. Ingersent, and S. Ulloa, Phys. Rev. Lett. {\bf 97}, 096603 (2006).


\bibitem{c8} M. Choi, R. Lopez, and R. Aguado, 
Phys. Rev. Lett. {\bf 95}, 067204 (2005).

\bibitem{c3} G. B. Martins, C. A. B\"usser,, K. A. Al-Hassanieh, A. Moreo, E. Dagotto, 
Phys. Rev. Lett. {\bf 94}, 026804 (2005).

\bibitem{silvestrov} P. G. Silvestrov and Y. Imry, cond-mat/0609355.

\bibitem{c10} K. Kikoin and Y. Oreg, cond-mat/0610207.

\bibitem{meden} V. Meden and F. Marquardt, Phys. Rev. Lett. {\bf 96}, 146801 (2006).

\bibitem{meden1} C. Karrasch, T. Hecht, A. Weichselbaum, J. von Delft, Y. Oreg, V. Meden, cond-mat/0612490.

\bibitem{wie} P. Wiegmann, V. Filyov, and A. Tsvelik, Sov. Phys. JETP Lett. 
{\bf 35} (1982) 77; A. Tsvelik and P. Wiegmann, Adv. in Phys. {\bf 32} (1983) 453.
N. Kawakami and A. Okiji, Phys. Lett. A {\bf 86}, 483 (1981).

\bibitem{long} R. M. Konik, H. Saleur, and A.W.W. Ludwig,
PRL {\bf 87}, 236801 (2001); ibid, PRB {\bf 66}, 125304 (2002).

\bibitem{short} R. M. Konik, cond-mat/0602617.

\bibitem{c6} 
P. Simon and D. Feinberg, Phys. Rev. Lett. 97, 247207 (2006).

\bibitem{kittel} C. Kittel, {\it Solid State Physics} (Academic Press, New York, 1968), Vol. 22.

\bibitem{utsumi} Y. Utsumi, J. Martinek, P. Bruno, and H. Imamura, Phys. Rev. B {\bf 69}, 155320 (2004).

\bibitem{fano} R. M. Konik, J. Stat. Mech. (2004) L11001; ibid, cond-mat/0401617.

\bibitem{andrei} N. Andrei, Phys. Lett. A {\bf 87}, 299 (1982).

\bibitem{halchoy} D. Haldane, T. Choy, Phys. Lett. A {\bf 90}, 83 (1981).

\bibitem{hubbard} T. Deguchi, F. Essler, F. G\"ohmann, A. Kl\"umper, V. Korepin, K. Kusakabe,
Physics Reports {\bf 331}, (2000) 197.

\bibitem{aus} S.Y. Cho, H.Q. Zhou and R.H. McKenzie, Phys. Rev. B {\bf 68}, 125327 (2003).

\bibitem{simon1} P. Simon and I Affleck, Phys. Rev. B {\bf 68}, 115304 (2003).

\bibitem{kawa_den} N. Kawakami and A. Okiji,  Phys. Rev. B {\bf 40} 7066 (1989). 

\bibitem{jones} B. A. Jones, B. G. Kotliar, and A. J. Millis, Phys. Rev. B. {\bf 39}, 3415 (1989).

\bibitem{haldane} D. Haldane, Phys. Rev. Lett. {\bf 40}, 416 (1978).

\bibitem{costi} T. Costi, A. Hewson, and V. Vlatic, J. Phys.: Cond. Mat. {\bf 6}, 2519 (1994).

\bibitem{langreth} D. Langreth, Phys. Rev. {\bf 150}, 516 (1966).

\bibitem{goldstein} M. Goldstein and R. Berkovits, cond-mat/0610810.

\bibitem{gefen} J. K\"onig, Y. Gefen, and G. Sch\"on, Phys. Rev. Lett. {\bf 81}, 4468 (1998).

\bibitem{kaminski} A. Kaminski, Y. Nazarov, and L. Glazman, Phys. Rev. B {\bf 62}, 8154 (2000).

\bibitem{hershfield} K. Majumdar, A. Schiller, and S. Hershfield, Phys. Rev. B {\bf 57}, 2991 (1998)

\bibitem{pustilnik} M. Pustilnik and L. Glazman, Phys. Rev. Lett. {\bf 87}, 216601 (2001).

\end{thebibliography}
\end{document}